\pdfoutput=1

\documentclass[11pt]{article}

\usepackage[preprint]{acl}

\usepackage{times}
\usepackage{latexsym}

\usepackage[T1]{fontenc}

\usepackage[utf8]{inputenc}

\usepackage{microtype}

\usepackage{inconsolata}

\usepackage{graphicx}
\usepackage{url}
\usepackage[most]{tcolorbox}
\tcbset{
  reviewbox/.style={
    enhanced,
    colback=gray!5,
    colframe=black, 
    boxrule=0.8pt,
    arc=1.5mm,
    auto outer arc,
    boxsep=5pt,
    left=6pt,
    right=6pt,
    top=6pt,
    bottom=6pt,
    fonttitle=\bfseries,
    before skip=10pt,
    after skip=10pt,
  }
}

\usepackage{multirow}
\usepackage{booktabs}
\usepackage{subcaption}
\usepackage{geometry}
\usepackage{graphicx}
\usepackage{listings}
\usepackage{xcolor}
\usepackage{hyperref}

\lstdefinestyle{paramstyle}{
    basicstyle=\ttfamily\small,
    backgroundcolor=\color{gray!10},
    frame=single,
    framerule=0.5pt,
    rulecolor=\color{gray!50},
    breaklines=true,
    showstringspaces=false,
    tabsize=2,
    xleftmargin=0.5cm,
    xrightmargin=0.5cm
}

\PassOptionsToPackage{table,xcdraw}{xcolor}

\usepackage{colortbl, booktabs}
\definecolor{lightgray}{gray}{0.9}
\definecolor{lightgreen}{RGB}{200, 255, 200}
\definecolor{lightred}{RGB}{255, 200, 200}
\definecolor{lightblue}{RGB}{230, 240, 255}
\definecolor{lightyellow}{RGB}{255, 250, 220}

\title{The Feasibility of Topic-Based Watermarking on Academic Peer Reviews}

\author{
  \textbf{Alexander Nemecek\textsuperscript{1}}, \,
  \textbf{Yuzhou Jiang\textsuperscript{1}}, \,
  \textbf{Erman Ayday\textsuperscript{1}}
\\
  \textsuperscript{1}Case Western Reserve University
\\
  \texttt{\{ajn98, yxj466, exa208\}@case.edu}
}

\begin{document}
\maketitle
\begin{abstract}
Large language models (LLMs) are increasingly integrated into academic workflows, with many conferences and journals permitting their use for tasks such as language refinement and literature summarization. However, their use in peer review remains prohibited due to concerns around confidentiality breaches, hallucinated content, and inconsistent evaluations. As LLM-generated text becomes more indistinguishable from human writing, there is a growing need for reliable attribution mechanisms to preserve the integrity of the review process. In this work, we evaluate topic-based watermarking (TBW), a semantic-aware technique designed to embed detectable signals into LLM-generated text. We conduct a systematic assessment across multiple LLM configurations, including base, few-shot, and fine-tuned variants, using authentic peer review data from academic conferences. Our results show that TBW maintains review quality relative to non-watermarked outputs, while demonstrating robust detection performance under paraphrasing. These findings highlight the viability of TBW as a minimally intrusive and practical solution for LLM attribution in peer review settings.\footnote{Our code and data is available at \url{https://github.com/ANCP2021/Watermarking-LLM-Conference}}

\end{abstract}

\section{Introduction}\label{intro}
As large language models (LLMs) continue to evolve, their adoption has accelerated in academic writing~\cite{dergaa2023human, editorials2023tools}. LLMs are widely used for language polishing, literature search, and low-novelty writing. Many conferences now explicitly allow authors to use LLMs for certain tasks, provided that authors retain full responsibility for the content~\cite{emnlp2025cfp, neurips2025llm, icml2025cfp}. These policies uphold pre-LLM expectations around authorship and accountability while adapting to new technology.

In contrast, the use of LLMs by peer reviewers is widely prohibited~\cite{emnlp2025review, neurips2025llm, icml2025review}. Such practices risk confidentiality breaches, low-quality evaluations, and data exposure~\cite{zhou-etal-2024-llm, maini2024llm}. Recent empirical studies suggest, however, that LLM-assisted reviews are already present in major conferences, leading to inflated scores, reduced reviewer confidence, and distortions in paper rankings~\cite{liang2024monitoring, latona2024ai, ye2024we}. These developments underscore the urgency of developing attribution mechanisms to detect and manage LLM usage in peer review.

Distinguishing between machine- and human-authored reviews has become difficult, as LLM-generated content continues to improve. This creates an urgent need for technical mechanisms to trace review provenance. \textit{Watermarking} offers a promising approach, embedding imperceptible, machine-detectable signatures into generated text~\cite{zhao2024sok}. However, existing work focuses on general-domain text, with limited analysis in peer review contexts~\cite{liu2024a, zhao2023provable}.

In this paper, we present the first systematic evaluation of topic-based watermarking (TBW) in the context of academic peer reviews, a high-stakes domain with distinct structural, semantic, and ethical constraints that differentiate it from general-purpose text generation settings. While TBW was originally proposed for open-domain text~\cite{nemecek2024topic}, its applicability to specialized domains with strict quality requirements and adversarial threat models has remained unexplored. We address this gap by adapting TBW to peer review workflows and conducting a comprehensive assessment of its feasibility as a practical attribution mechanism for conference organizers.

\textbf{Our contribution} is threefold. First, we demonstrate that peer review represents an ideal use case for TBW due to its inherent structural properties: reviews must remain topically aligned with the paper under evaluation, satisfying TBW's topic-matching assumption naturally, unlike open-ended generation tasks where topic drift undermines detection. Second, we introduce domain-specific adaptations including custom topic sets (\{theory, applications, models, optimization\}) aligned with machine learning conference reviews, and we systematically evaluate TBW across three model configurations: base, few-shot, and fine-tuned, representing varying levels of model adaptation. Third, we assess robustness under realistic paraphrasing threats (PEGASUS and DIPPER), reflecting scenarios where reviewers may rephrase generated content to improve clarity or evade detection while preserving review quality.

Without effective attribution mechanisms, the credibility and rigor of academic conferences could erode, leading to lower-quality evaluations and increased reliance on potentially unverifiable, machine-generated feedback. Watermarking provides a practical and minimally disruptive approach for LLM accountability, helping to safeguard academic standards while accommodating the evolving role of generative models. Related work on LLM watermarking and detection approaches is provided in \S\ref{prelim}.

\section{Topic-Based Watermarks}\label{methods}
Topic-based watermarking (TBW)~\cite{nemecek2024topic} is a semantic-aware watermark that subtly influences a language model's token selection process to leave a detectable signature. Unlike earlier schemes such as KGW~\cite{kirchenbauer2023watermark}, which rely on randomly partitioned vocabularies using a secret seed, TBW constructs topic-specific token subsets (``green lists'') aligned with the semantic content of the input prompt through a deterministic, keyless mapping. This approach helps preserve fluency while enhancing robustness against paraphrasing.

\noindent\textbf{Token-to-Topic Mappings.} TBW assigns tokens to topic-specific green lists using semantic similarity without requiring a master key or secret seed. A small set of topics ${t_1, \dots, t_K}$ is defined, each represented by an embedding $\mathbf{e}_{t_i}$ computed via a sentence embedding model. Each token $v \in V$ in the model's vocabulary is embedded as $\mathbf{e}_v$, and its cosine similarity with each topic embedding is computed. If the maximum similarity exceeds a threshold $\tau$, the token is assigned to the green list $G_{t_i}$ for the most similar topic. Tokens that do not meet this threshold are placed in a residual set and evenly distributed across all green lists to maintain full vocabulary coverage. This deterministic assignment process is reproducible given only the embedding model, topic set, and threshold $\tau$ as no secret parameters are involved.\footnote{While this keyless design is suitable for practical threats like paraphrasing, TBW can be extended with a keyed residual assignment (e.g., using a secret seed to distribute non-matching tokens) if cryptographically stronger adversaries are of concern.}

\noindent\textbf{Generation \& Detection.} During generation, the most relevant topic is identified from the input prompt using keyword extraction, and TBW adds a small logit bias $\delta$ to all tokens in the corresponding green list. This increases the likelihood of sampling topic-aligned tokens while preserving the model architecture and generation efficiency. The watermark strength is controlled by $\delta$: higher values produce stronger attribution signals but may cause detectable shifts in word choice.

For detection, TBW recovers the relevant topic from the input text and counts green-list tokens $g$ relative to total tokens $n$. A $z$-score quantifies whether the green-token rate exceeds an expected baseline proportion $\gamma$:
$$
z \;=\; \frac{g \;-\; \gamma \;\cdot\; n}{\sqrt{n \;\cdot\; \gamma \;\cdot\; (1 - \gamma)}}.
$$
If $z > z_{\text{threshold}}$, the text is classified as watermarked.

\noindent\textbf{Rationale for TBW.} We select TBW for its combination of robustness, adaptability, and minimal overhead~\cite{nemecek2024topic}. TBW is particularly well-suited to peer review contexts where paraphrasing represents a realistic threat model where reviewers may rephrase generated content to improve clarity or avoid detection, but are unlikely to introduce noise that would degrade quality. 

Additionally, TBW's requirement of topic alignment between the input prompt and generated text naturally holds in peer review, where content must remain semantically aligned with the paper under evaluation. This structural constraint provides an advantage over general-domain watermarking because the detector (conference organizer) has direct access to the same paper submission used at generation time, they can reliably recover the correct topic and reconstruct the identical green lists without requiring any secret key. This eliminates key dependency entirely, the deterministic token-to-topic mapping combined with guaranteed topic alignment means detection requires only the public submission text, not hidden parameters.   

Finally, TBW's semantic biasing strategy preserves fluency and style while supporting domain adaptation through customizable topic sets  ${t_1, \dots, t_K}$ that can be tailored to specific research fields or venues.

\section{Experimental Setup}\label{genTask}
We simulate realistic LLM-based peer review generation by training and prompting models to write reviews conditioned on a paper's title and abstract. We condition on paper abstracts rather than full text due to context length constraints and structured data availability.

\subsection{Dataset} 
We construct a dataset of paper titles, abstracts, and corresponding reviews from ICLR and NeurIPS conferences using the OpenReview API~\cite{openreview-api}. To minimize risk of including LLM-generated content, we restrict data to conferences before ChatGPT's release (November 2022)~\cite{openai_chatgpt_2022}: ICLR 2018-2023 and NeurIPS 2021-2022. The final dataset contains approximately 19,000 reviews, each including summary, strengths/weaknesses, and recommendation scores. For each paper, we randomly sample a single review to construct prompt-completion pairs, ensuring reviewer diversity while avoiding overrepresentation. Detailed statistics are in \text{\S}\ref{appendA1}.

\subsection{TBW Domain Adaptation}\label{config}
We adapt TBW to peer review by modifying the topic sets while retaining original parameter settings. Instead of general-purpose topics (e.g., \texttt{technology}, \texttt{sports}), we define domain-specific topics: \{\texttt{theory}, \texttt{applications}, \texttt{models}, \texttt{optimization}\} to capture themes in ML conference reviews. We use logit bias $\delta = 2.0$ and similarity threshold $\tau = 0.7$ for primary experiments, with additional evaluation at $\tau = 0.3$. Complete parameter details are provided in \text{\S}\ref{watermarking_params_details}.

\subsection{Model Configurations} 
To assess the feasibility of TBW across varying levels of model adaptation and reviewer effort, we utilize the pretrained \texttt{Llama-3.1-8B} base checkpoint~\cite{grattafiori2024llama} in three configurations: base, few-shot, and fine-tuned. 
The base configuration uses the pretrained model checkpoint directly without additional training or prompt engineering, simulating minimal reviewer effort. The few-shot setting provides the same base model with example peer reviews as part of the input prompt, enabling it to replicate the expected format and tone with lightweight guidance. The fine-tuned configuration involves supervised training on peer review data using parameter-efficient methods, resulting in a model more aligned with the review-writing task. This model size offers a practical balance between computational efficiency and generation quality suitable for our multi-configuration experiments. 
Detailed setup parameters for few-shot prompting and fine-tuning are provided in \text{\S}\ref{app:model_setup}.

\section{Experiments}\label{exp}
We conduct a series of experiments across multiple dimensions, including text quality, robustness to paraphrasing, and classifier-based attribution.

\subsection{Generation Quality}\label{gen_qual}
To assess TBW's impact on generation quality, we evaluate perplexity across 1,000 samples per model configuration (base, few-shot, fine-tuned), comparing against unwatermarked baselines and two existing schemes: KGW~\cite{kirchenbauer2023watermark} and Google's SynthID-Text~\cite{dathathri2024scalable}. Each sample contains approximately $200 \pm 5$ tokens. Parameter configurations, additional threshold results, human evaluations, and BERTScore evaluations are detailed in \text{\S}\ref{appendixD}.

\begin{figure}[h]
\begin{center}
\centerline{\includegraphics[width=1\columnwidth]{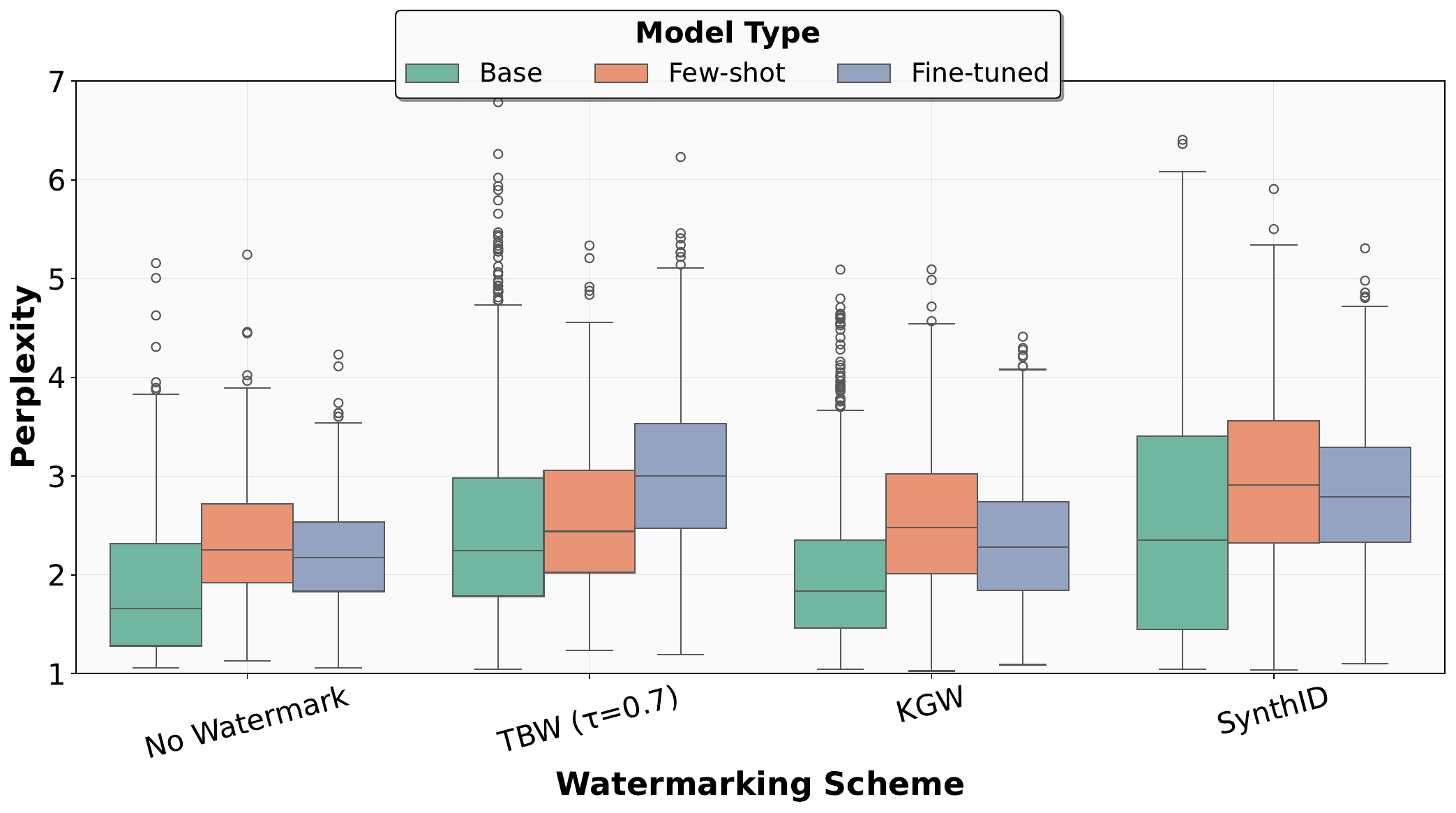}}
\caption{Perplexity distributions across model and watermark configurations. Lower values indicate better fluency. Values above 20 are truncated for clarity.}
\label{perplexity-07}
\end{center}
\end{figure}

We compute perplexity using the generating model (\texttt{Llama-3.1-8B}) as a fluency proxy, where lower values indicate higher naturalness. For visualization clarity, values above 20 are truncated in Figure~\ref{perplexity-07}.
\noindent The results demonstrate that TBW introduces minimal perplexity degradation while maintaining fluency compared to existing watermarking schemes. While Figure~\ref{perplexity-07} shows TBW with slightly higher median perplexity than KGW in some configurations, the truncation at perplexity 20 obscures an important quality difference. Table~\ref{tab:retention} reports the number of samples retained below this threshold across all watermarking schemes and model configurations.

\begin{table}[h!]
\centering
\begin{tabular}{llc}
\toprule
\textbf{Model} & \textbf{Scheme} & \textbf{Samples Retained} \\
\midrule
\multirow{4}{*}{Base} & NW & 508 \\
                      & TBW & \textbf{991} \\
                      & KGW & 840 \\
                      & SynthID & 538 \\
\cmidrule(){1-3}
Few-shot & All schemes & \textbf{1000} \\
\cmidrule(){1-3}
Fine-tuned & All schemes & \textbf{1000} \\
\bottomrule
\end{tabular}
\caption{Number of retained generations with perplexity $\leq 20$ across model configurations. TBW retains 991/1000 base model samples, significantly outperforming KGW (840/1000) and SynthID (538/1000), demonstrating superior outlier suppression.}
\label{tab:retention}
\end{table}
\noindent Notably, TBW retained 991 of 1000 base model samples below the perplexity threshold, excluding only 9 outliers, whereas KGW retained 840 (160 excluded) and SynthID retained only 538 (462 excluded). This demonstrates that TBW produces far fewer low-quality outliers than existing methods, the best performance among all watermarking schemes tested. These results suggest TBW preserves generation naturalness but enhances lexical consistency in low-context scenarios by steering generation toward topic-relevant vocabulary.

While perplexity serves as a useful fluency proxy, it does not fully capture semantic fidelity to the original content. To complement this analysis, we evaluate semantic similarity using BERTScore~\cite{zhang2019bertscore}, which measures contextual embedding alignment between generated reviews and ground-truth references (\text{\S}\ref{bertbert}), and conduct a small-scale human evaluation on the fluency, coherence, and usefulness of generated reviews (\text{\S}\ref{humanevals}). Furthermore, we evaluate TBW under a relaxed similarity threshold ($\tau = 0.3$) to assess sensitivity to green-list construction (\text{\S}\ref{lowerbertbert}). These complementary metrics confirm that TBW preserves both fluency and semantic integrity across generation conditions.

\subsection{Robustness to Paraphrasing Attacks}\label{robust}
We evaluate TBW's resilience against paraphrasing attacks, which represent a realistic threat model where reviewers may rephrase LLM-generated content to evade detection while preserving semantic meaning. We focus on full-text paraphrasing attacks that best reflect plausible reviewer behavior, excluding token-level or partial edit scenarios.

\begin{table}[t]
\centering
\small
\begin{tabular}{clccc}
\toprule
 & & \multicolumn{3}{c}{\textbf{Detection Accurcy}} \\
\textbf{Model} & \textbf{Attacks} & \textbf{TBW} & KGW & SynthID \\
\midrule
\multirow{3}{*}{Base}
  & No Attack &  0.946 & \textbf{0.971} & 0.909 \\
  & PEGASUS  & \textbf{0.847} & 0.477 & 0.135 \\
  & DIPPER   & \textbf{0.876} & 0.754 & 0.173 \\
\midrule
\multirow{3}{*}{Few-shot}
  & No Attack & 0.622 & \textbf{0.975} & 0.959 \\
  & PEGASUS  & \textbf{0.580} & \textbf{0.580} & 0.359 \\
  & DIPPER   & 0.517 & \textbf{0.748} & 0.225 \\
\midrule
\multirow{3}{*}{Fine-tuned}
  & No Attack & 0.880 & 0.926 & \textbf{0.960} \\
  & PEGASUS  & \textbf{0.583} & 0.437 & 0.180\\
  & DIPPER   & 0.584 & \textbf{0.657} & 0.159 \\
\bottomrule
\end{tabular}
\caption{Detection accuracy of TBW, KGW, and SynthID across model configurations and paraphrasing attack types. Scores reflect the proportion of correctly identified watermarked samples per condition.}
\label{tpr_fpr_results}
\end{table}

We generate 1,000 samples per model configuration (base, few-shot, fine-tuned) for each watermarking scheme, with each sample containing $200 \pm 5$ tokens. Paraphrasing attacks are implemented using two established models: PEGASUS~\cite{zhang2020pegasus} and DIPPER~\cite{krishna2023paraphrasing} (configured with \texttt{lexical=60} and \texttt{order=40}), following standard robustness evaluation protocols~\cite{hou-etal-2024-semstamp, liu2024adaptive}.

Table~\ref{tpr_fpr_results} shows detection accuracy across three attack conditions. TBW demonstrates superior robustness compared to existing methods, maintaining detection accuracy for base models and fine-tuned models even under paraphrasing attacks. In contrast, KGW and SynthID show degradation, with SynthID dropping to as low as 13.5\% accuracy under PEGASUS attacks. 

The few-shot configuration exhibits reduced performance across all methods, with TBW achieving 58\% accuracy under PEGASUS and 52\% under DIPPER. This degradation likely stems from topic misalignment between the prompt examples and target papers, which weakens topic alignment and reduces detectability post-paraphrasing. 

Importantly, this represents a characteristic trade-off of TBW's semantic design. While topic-aware watermarking provides robustness in controlled settings (base and fine-tuned models), it becomes sensitive to exemplar-target mismatch in few-shot prompting. This sensitivity surfaces an important consideration for effective deployment of TBW in few-shot scenarios, which would benefit from improved exemplar selection strategies or dynamic prompt construction that ensures topical consistency between examples and the target generation task. 

We verify that TBW maintains low false positive rates on human-written reviews through its vocabulary partitioning strategy, which preserves lexical diversity across topic-specific green lists. Complete ROC analysis is provided in \text{\S}\ref{appendE}.

\subsection{Classifier-Based Attribution}\label{class-based}
Beyond detecting the presence of watermarks, one question is whether watermarking degrades the semantic integrity of generated reviews in ways that affect downstream interpretation tasks. Specifically, if watermarking distorts the linguistic features that correlate with review sentiment (accept, borderline, reject), it could undermine the very purpose of generating coherent and evaluable reviews.

To address this concern, we evaluate whether both watermarked and non-watermarked LLM-generated peer reviews can be accurately attributed to their original review labels (accept, borderline, reject) using transformer-based classification models. This experiment serves two purposes: (i) it validates that watermarked reviews preserve sufficient semantic structure to support rating prediction, and (ii) it tests whether TBW's topic-aware biasing introduces artifacts that either help or hinder sentiment classification. If watermarking significantly degrades classifier performance, it would suggest that the semantic alterations required for attribution compromise review quality in measurable ways.

\begin{table}[t]
\centering
\scriptsize
\begin{tabular}{@{}l|cc|cc|cc@{}}
\toprule
& \multicolumn{2}{c|}{\textbf{Base}} & \multicolumn{2}{c|}{\textbf{Few-shot}} & \multicolumn{2}{c}{\textbf{Fine-tuned}} \\
\textbf{Metric} & \textbf{NW} & \textbf{TBW} & \textbf{NW} & \textbf{TBW} & \textbf{NW} & \textbf{TBW} \\
\midrule
\multicolumn{7}{c}{\textit{BERT}} \\
\midrule
Accuracy & 0.290 & \textbf{0.321} & 0.403 & \textbf{0.437} & 0.400 & \textbf{0.416} \\
Precision & \textbf{0.353} & 0.346 & \textbf{0.373} & 0.366 & \textbf{0.367} & 0.366 \\
Recall & 0.328 & \textbf{0.342} & \textbf{0.379} & 0.369 & \textbf{0.370} & 0.367 \\
F1 & 0.278 & \textbf{0.317} & \textbf{0.360} & 0.358 & 0.364 & \textbf{0.366} \\
\midrule
\multicolumn{7}{c}{\textit{RoBERTa}} \\
\midrule
Accuracy & \textbf{0.486} & 0.432 & 0.399 & \textbf{0.424} & 0.406 & \textbf{0.443} \\
Precision & 0.344 & \textbf{0.357} & 0.362 & \textbf{0.371} & 0.367 & \textbf{0.401} \\
Recall & 0.341 & \textbf{0.352} & 0.368 & \textbf{0.371} & 0.374 & \textbf{0.403} \\
F1 & 0.305 & \textbf{0.350} & 0.337 & \textbf{0.353} & 0.367 & \textbf{0.402} \\
\bottomrule
\end{tabular}
\caption{Overall classification performance on original LLM-generated reviews. Metrics are averaged over Accept, Borderline, and Reject classes.}
\label{tab:overall-performance}
\end{table}

We use a balanced dataset of 9,000 training samples (3,000 per class) and 1,000 test samples. Reviews are mapped to three categories based on original scores: 1-4 (reject), 5-6 (borderline), and 7-10 (accept). We train BERT~\cite{devlin-etal-2019-bert} and RoBERTa~\cite{zhuang-etal-2021-robustly} classifiers to predict review ratings based on generated review text, evaluating performance on both watermarked and non-watermarked reviews.

As shown in Table~\ref{tab:overall-performance}, TBW demonstrates mixed but generally positive effects on classification performance. While some configurations show modest degradation (e.g., RoBERTa base accuracy), the majority of results indicate that TBW causes little to no degradation and often leads to improvements in both accuracy and F1 scores. We hypothesize that this occurs because topic-based watermarking encourages more topically consistent language that aligns better with the underlying review content.

These findings reinforce TBW's suitability for domain-sensitive contexts like peer review, where both traceability and semantic fidelity are critical. The results demonstrate that watermarking does not compromise and may even enhance the interpretability of generated reviews with respect to their underlying evaluative stance. Additional experimental details, including class-specific performance metrics, evaluation under lower similarity thresholds ($\tau = 0.3$), and robustness evaluation under paraphrasing attacks, are provided in \text{\S}\ref{classifier-training}.

\section{Conclusion}\label{con}
We present the first systematic evaluation of topic-based watermarking (TBW) for academic peer review, demonstrating that TBW preserves generation quality while maintaining robust detection under paraphrasing attacks. Our experiments across base, few-shot, and fine-tuned LLM configurations show that TBW's semantic grounding naturally aligns with peer review constraints, where content must remain topically consistent with the evaluated paper. These findings highlight TBW's viability as a minimally intrusive solution for LLM attribution in peer review settings, offering a practical mechanism to safeguard academic evaluation integrity. Discussion of deployment considerations and limitations is provided in \text{\S}\ref{diss}.

\newpage

\section*{Limitations}
This work inherits a key limitation of topic-based watermarking: the topic-matching assumption. As noted in the original proposal~\cite{nemecek2024topic}, watermark detection may degrade if the semantic topic of the generated output drifts significantly from the original prompt. This is particularly challenging in open-domain generation, where the input prompt is often unavailable at detection time. However, in the context of peer review, this limitation is largely mitigated. Reviewers must prompt the LLM using the content of the paper, either by directly including the text or referencing its abstract and title, ensuring that the generated review remains topically aligned with the source. Furthermore, during detection, conference organizers have access to the submission itself, allowing them to reliably identify the intended topic and recover the correct green list. As a result, the topic-matching assumption holds in this use case.

A second limitation concerns deployment and coverage. For watermarking to serve as a reliable attribution mechanism, it must be consistently applied across all LLMs used in a given environment. This is a general challenge for watermarking approaches and not unique to TBW. If only certain LLM providers implement watermarking while others do not, users can simply switch to unwatermarked systems to bypass attribution. While the governance and policy mechanisms required to address this challenge are beyond the scope of this paper, we acknowledge that the effectiveness of TBW in real-world enforcement depends on broader coordination across providers and platforms.

Thirdly, our evaluation of TBW's robustness in the context of academic peer reviews focuses on full-text paraphrasing attacks, which represent strong adversarial models where the entire review is rephrased. We selected this threat model because it represents the most comprehensive transformation: if TBW remains detectable when all text is paraphrased, it should naturally be robust to partial edits affecting only portions of the review. However, we acknowledge that real-world reviewer behavior may include additional manipulation strategies not evaluated in this work, such as hybrid human-LLM composition (mixing original writing with generated content), section reordering, selective compression or expansion of specific review components, and multi-stage translation attacks (e.g., English to an intermediate language and then back to English). While these scenarios represent valuable directions for future robustness analysis, we expect TBW's semantic grounding to provide resilience against such attacks, particularly in cases where the topical alignment between the paper and review is preserved. Future work should systematically evaluate these additional threat models to fully characterize TBW's operational boundaries.

An important distinction not addressed in our current evaluation is differentiating between text originally authored by a human reviewer but subsequently polished by an LLM (e.g., grammar correction, style refinement) versus text entirely generated by an LLM. This represents a challenging but critical use case, as many conferences permit LLM-assisted writing improvement while prohibiting wholesale generation. TBW's detection mechanism, which relies on green-list token density, would likely flag both scenarios similarly since the polishing process may still introduce watermarked tokens. Developing attribution methods that can reliably distinguish between these two cases remains an important direction for future work and may require complementary techniques beyond watermarking alone.

\section*{Ethical Considerations}
This work addresses the growing concern of unauthorized LLM usage in academic peer review. While many conferences permit LLM use for authoring papers, they explicitly prohibit it for generating reviews, citing risks to confidentiality, fairness, and accountability. Our goal is not to penalize reviewers but to support conference organizers in enforcing existing policies through lightweight and interpretable attribution tools. Topic-based watermarking introduces no additional risk to authors or reviewers, as it operates at the generation level without modifying model internals or relying on invasive detection mechanisms. We advocate for transparent disclosure of LLM usage in reviews and emphasize that attribution tools should be deployed with clear governance structures and ethical oversight.

\section*{Acknowledgments}
This work was partially supported by the National Science Foundation (NSF) under grant numbers 2050410 and OAC-2112606. ChatGPT and Claude were used in the preparation of this work for editing and grammar checking. No passages were copied without full author review, and all substantive ideas, analysis, and conclusions are the product and responsibility of the authors. Additionally, the listed AI tools were utilized for code development and early manuscript reviews.

\bibliography{anthology,custom}

@article{dergaa2023human,
  title={From human writing to artificial intelligence generated text: examining the prospects and potential threats of ChatGPT in academic writing},
  author={Dergaa, Ismail and Chamari, Karim and Zmijewski, Piotr and Saad, Helmi Ben},
  journal={Biology of sport},
  volume={40},
  number={2},
  pages={615--622},
  year={2023},
  publisher={Termedia}
}

@article{editorials2023tools,
  title={Tools such as ChatGPT threaten transparent science; here are our ground rules for their use},
  author={Editorials, Nature},
  journal={Nature},
  volume={613},
  number={7945},
  pages={612},
  year={2023}
}

@misc{icml2025cfp,
  author       = {{ICML}},
  title        = {ICML 2025 Call for Papers},
  howpublished = {\url{https://icml.cc/Conferences/2025/CallForPapers}},
  note         = {Accessed: 2025-05-15},
  year         = {2025}
}

@misc{emnlp2025cfp,
  author       = {{ACL}},
  title        = {ACL Rolling Review Call For Papers},
  howpublished = {\url{https://aclrollingreview.org/cfp#long-papers}},
  note         = {Accessed: 2025-05-15},
  year         = {2025}
}

@misc{neurips2025llm,
  author       = {{NeurIPS}},
  title        = {NeurIPS 2025 Policy on the Use of Large Language Models},
  howpublished = {\url{https://neurips.cc/Conferences/2025/LLM}},
  note         = {Accessed: 2025-05-15},
  year         = {2025}
}

@misc{icml2025review,
  author       = {{ICML}},
  title        = {ICML 2025 Reviewer Instructions},
  howpublished = {\url{https://icml.cc/Conferences/2025/ReviewerInstructions}},
  note         = {Accessed: 2025-05-15},
  year         = {2025}
}

@misc{emnlp2025review,
  author       = {{ACL}},
  title        = {ARR Reviewer Guidelines},
  howpublished = {\url{https://aclrollingreview.org/reviewerguidelines}},
  note         = {Accessed: 2025-05-15},
  year         = {2025}
}

@article{maini2024llm,
  title={LLM Dataset Inference: Did you train on my dataset?},
  author={Maini, Pratyush and Jia, Hengrui and Papernot, Nicolas and Dziedzic, Adam},
  journal={Advances in Neural Information Processing Systems},
  volume={37},
  pages={124069--124092},
  year={2024}
}

@article{nemecek2024topic,
  title={Topic-Based Watermarks for Large Language Models},
  author={Nemecek, Alexander and Jiang, Yuzhou and Ayday, Erman},
  journal={arXiv preprint arXiv:2404.02138},
  year={2024}
}

@inproceedings{kirchenbauer2023watermark,
  title={A watermark for large language models},
  author={Kirchenbauer, John and Geiping, Jonas and Wen, Yuxin and Katz, Jonathan and Miers, Ian and Goldstein, Tom},
  booktitle={International Conference on Machine Learning},
  pages={17061--17084},
  year={2023},
  organization={PMLR}
}

@article{zhao2024sok,
  title={SoK: Watermarking for AI-Generated Content},
  author={Zhao, Xuandong and Gunn, Sam and Christ, Miranda and Fairoze, Jaiden and Fabrega, Andres and Carlini, Nicholas and Garg, Sanjam and Hong, Sanghyun and Nasr, Milad and Tramer, Florian and others},
  journal={arXiv preprint arXiv:2411.18479},
  year={2024}
}

@article{dathathri2024scalable,
  title={Scalable watermarking for identifying large language model outputs},
  author={Dathathri, Sumanth and See, Abigail and Ghaisas, Sumedh and Huang, Po-Sen and McAdam, Rob and Welbl, Johannes and Bachani, Vandana and Kaskasoli, Alex and Stanforth, Robert and Matejovicova, Tatiana and Hayes, Jamie and Vyas, Nidhi and Merey, Majd Al and Brown-Cohen, Jonah and Bunel, Rudy and Balle, Borja and Cemgil, Taylan and Ahmed, Zahra and Stacpoole, Kitty and Shumailov, Ilia and Baetu, Ciprian and Gowal, Sven and Hassabis, Demis and Kohli, Pushmeets},
  journal={Nature},
  volume={634},
  number={8035},
  pages={818--823},
  year={2024},
  publisher={Nature Publishing Group UK London}
}

@article{zhao2023provable,
  title={Provable robust watermarking for ai-generated text},
  author={Zhao, Xuandong and Ananth, Prabhanjan and Li, Lei and Wang, Yu-Xiang},
  journal={arXiv preprint arXiv:2306.17439},
  year={2023}
}

@inproceedings{liu2024a,
title={A Semantic Invariant Robust Watermark for Large Language Models},
author={Aiwei Liu and Leyi Pan and Xuming Hu and Shiao Meng and Lijie Wen},
booktitle={The Twelfth International Conference on Learning Representations},
year={2024},
url={https://openreview.net/forum?id=6p8lpe4MNf}
}

@article{li2024use,
  title={Use of Artificial Intelligence in Peer Review Among Top 100 Medical Journals},
  author={Li, Zhi-Qiang and Xu, Hui-Lin and Cao, Hui-Juan and Liu, Zhao-Lan and Fei, Yu-Tong and Liu, Jian-Ping},
  journal={JAMA Network Open},
  volume={7},
  number={12},
  pages={e2448609--e2448609},
  year={2024},
  publisher={American Medical Association}
}

@article{liang2024monitoring,
  title={Monitoring ai-modified content at scale: A case study on the impact of chatgpt on ai conference peer reviews},
  author={Liang, Weixin and Izzo, Zachary and Zhang, Yaohui and Lepp, Haley and Cao, Hancheng and Zhao, Xuandong and Chen, Lingjiao and Ye, Haotian and Liu, Sheng and Huang, Zhi and others},
  journal={arXiv preprint arXiv:2403.07183},
  year={2024}
}

@article{ye2024we,
  title={Are we there yet? revealing the risks of utilizing large language models in scholarly peer review},
  author={Ye, Rui and Pang, Xianghe and Chai, Jingyi and Chen, Jiaao and Yin, Zhenfei and Xiang, Zhen and Dong, Xiaowen and Shao, Jing and Chen, Siheng},
  journal={arXiv preprint arXiv:2412.01708},
  year={2024}
}

@article{latona2024ai,
  title={The ai review lottery: Widespread ai-assisted peer reviews boost paper scores and acceptance rates},
  author={Latona, Giuseppe Russo and Ribeiro, Manoel Horta and Davidson, Tim R and Veselovsky, Veniamin and West, Robert},
  journal={arXiv preprint arXiv:2405.02150},
  year={2024}
}

@article{yu2025your,
  title={Is Your Paper Being Reviewed by an LLM? A New Benchmark Dataset and Approach for Detecting AI Text in Peer Review},
  author={Yu, Sungduk and Luo, Man and Madusu, Avinash and Lal, Vasudev and Howard, Phillip},
  journal={arXiv preprint arXiv:2502.19614},
  year={2025}
}

@inproceedings{kumar2025mixrevdetect,
  title={MixRevDetect: Towards Detecting AI-Generated Content in Hybrid Peer Reviews.},
  author={Kumar, Sandeep and Garg, Samarth and Sengupta, Sagnik and Ghosal, Tirthankar and Ekbal, Asif},
  booktitle={Proceedings of the 2025 Conference of the Nations of the Americas Chapter of the Association for Computational Linguistics: Human Language Technologies (Volume 2: Short Papers)},
  pages={944--953},
  year={2025}
}

@article{rao2025detecting,
  title={Detecting LLM-Written Peer Reviews},
  author={Rao, Vishisht and Kumar, Aounon and Lakkaraju, Himabindu and Shah, Nihar B},
  journal={arXiv preprint arXiv:2503.15772},
  year={2025}
}

@misc{openreview-api,
  author       = {OpenReview},
  title        = {OpenReview Documentation},
  year         = {2024},
  howpublished = {\url{https://docs.openreview.net/getting-started/using-the-api}},
  note         = {Accessed: 2025-05-16}
}

@misc{openai_chatgpt_2022,
  title        = {Introducing ChatGPT},
  author       = {{OpenAI}},
  year         = {2022},
  month        = {November},
  url          = {https://openai.com/index/chatgpt/},
  note         = {Accessed: 2025-05-16}
}

@article{grattafiori2024llama,
  title={The llama 3 herd of models},
  author={Grattafiori, Aaron and Dubey, Abhimanyu and Jauhri, Abhinav and Pandey, Abhinav and Kadian, Abhishek and Al-Dahle, Ahmad and Letman, Aiesha and Mathur, Akhil and Schelten, Alan and Vaughan, Alex and others},
  journal={arXiv preprint arXiv:2407.21783},
  year={2024}
}

@inproceedings{reimers-2020-multilingual-sentence-bert,
    title = "Making Monolingual Sentence Embeddings Multilingual using Knowledge Distillation",
    author = "Reimers, Nils and Gurevych, Iryna",
    booktitle = "Proceedings of the 2020 Conference on Empirical Methods in Natural Language Processing",
    month = "11",
    year = "2020",
    publisher = "Association for Computational Linguistics",
    url = "https://arxiv.org/abs/2004.09813",
}

@misc{grootendorst2020keybert,
  author       = {Maarten Grootendorst},
  title        = {KeyBERT: Minimal keyword extraction with BERT.},
  year         = 2020,
  publisher    = {Zenodo},
  version      = {v0.3.0},
  doi          = {10.5281/zenodo.4461265},
  url          = {https://doi.org/10.5281/zenodo.4461265}
}

@article{zhang2019bertscore,
  title={Bertscore: Evaluating text generation with bert},
  author={Zhang, Tianyi and Kishore, Varsha and Wu, Felix and Weinberger, Kilian Q and Artzi, Yoav},
  journal={arXiv preprint arXiv:1904.09675},
  year={2019}
}

@article{liu2024adaptive,
  title={Adaptive text watermark for large language models},
  author={Liu, Yepeng and Bu, Yuheng},
  journal={arXiv preprint arXiv:2401.13927},
  year={2024}
}

@article{roberta,
  author    = {Yinhan Liu and
               Myle Ott and
               Naman Goyal and
               Jingfei Du and
               Mandar Joshi and
               Danqi Chen and
               Omer Levy and
               Mike Lewis and
               Luke Zettlemoyer and
               Veselin Stoyanov},
  title     = {RoBERTa: {A} Robustly Optimized {BERT} Pretraining Approach},
  journal   = {CoRR},
  volume    = {abs/1907.11692},
  year      = {2019},
  url       = {http://arxiv.org/abs/1907.11692},
  archivePrefix = {arXiv},
  eprint    = {1907.11692},
  bibsource = {dblp computer science bibliography, https://dblp.org}
}

@inproceedings{zhang2020pegasus,
  title={Pegasus: Pre-training with extracted gap-sentences for abstractive summarization},
  author={Zhang, Jingqing and Zhao, Yao and Saleh, Mohammad and Liu, Peter},
  booktitle={International conference on machine learning},
  pages={11328--11339},
  year={2020},
  organization={PMLR}
}

@article{krishna2023paraphrasing,
  title={Paraphrasing evades detectors of ai-generated text, but retrieval is an effective defense},
  author={Krishna, Kalpesh and Song, Yixiao and Karpinska, Marzena and Wieting, John and Iyyer, Mohit},
  journal={Advances in Neural Information Processing Systems},
  volume={36},
  pages={27469--27500},
  year={2023}
}


\appendix

\section{Related Work}\label{prelim}
Since the release of ChatGPT, LLMs have been rapidly adopted across various stages of the academic workflow. Their use has raised concerns about authorship and peer review integrity. Most conferences and journals now permit authors to leverage LLMs; however, this permissive stance does not extend to peer reviewers. Leading venues such as NeurIPS and ACL explicitly prohibit the use of LLMs by reviewers~\cite{neurips2025llm, emnlp2025review}. These policies reflect growing concerns around review quality, including the risk of shallow or hallucinated feedback, reduced technical depth, and breaches of confidentiality that would compromise the double-blind review process~\cite{li2024use}.

Despite these restrictions, recent studies suggest that LLM-assisted reviews are already present at major conferences. \citet{liang2024monitoring} estimate that 5-15\% of reviews were substantially modified using LLMs, with affected reviewers showing lower confidence and less engagement during rebuttals. \citet{latona2024ai} report similar trends and observe a score inflation effect, while \citet{ye2024we} show that even subtle LLM manipulations can shift paper rankings. Together, these findings underscore the risks unauthorized LLM use poses to peer review fairness and rigor.

Given the increasing use of LLMs for peer review generation, recent work has focused on detecting and attributing such content. Much of this research explores classifier-based detection or semantic similarity methods aimed at identifying AI-generated text. For example, \citet{yu2025your} propose a detection method based on the semantic similarity between a known LLM-generated review and a test review, flagging a review as machine-generated when similarity exceeds a threshold. Similarly, \citet{kumar2025mixrevdetect} introduce a partition-based method under the assumption that a review contains both human- and LLM-written components. They segment the review into distinct points, complete each segment with a reference LLM, and measure semantic similarity between these completions and the original text to detect potential LLM involvement.

However, these detection methods fail under paraphrasing or hybrid-review scenarios, where even minor edits or partial human rewriting can evade detection. To address this limitation, watermarking offers a promising alternative by embedding identifiable signals directly into the generated text. One foundational method is the KGW algorithm~\cite{kirchenbauer2023watermark}, which partitions the model's vocabulary into ``green'' and ``red'' token sets. During generation, the model is subtly biased to sample more frequently from the ``green'' list, which acts as a watermark-carrying set, while avoiding tokens in the ``red'' list. This results in output text that biases outputs toward ``green'' tokens with minimal quality loss. Variants aim to improve robustness and preserve quality~\cite{liu2024a, zhao2023provable, hou-etal-2024-semstamp}.

More recently, commercial systems have also entered this space. For example, Google's SynthID-Text watermarking system employs a strategy called Tournament Sampling, in which candidate tokens are ranked according to randomized watermarking functions, and the highest-ranked token is selected during generation~\cite{dathathri2024scalable}. While both academic and commercial watermarking approaches have shown promise, they are primarily evaluated on general-purpose domains such as news or encyclopedic text, and rarely tested under the stylistic and ethical constraints found in peer review.

While a few frameworks target peer review watermarking~\cite{rao2025detecting}, they rely on tightly integrated pipelines and lack evaluation across adaptation modes. Topic-based watermarking (TBW)~\cite{nemecek2024topic}, originally proposed for open-domain text, provides a lightweight, semantically guided alternative.

\section{Discussion}\label{diss}
Topic-based watermarking performs particularly well in the peer review setting due to the natural alignment between the subject of a paper and the content of its corresponding review. Unlike more open-ended generation tasks, peer reviews are tightly grounded in the paper being evaluated, making significant topic shifts unlikely, unless introduced deliberately by the reviewer. Since high-quality, relevant reviews are needed for the academic evaluation process, such intentional degradation is improbable in practice.

We also observe that topic-based watermarking is compatible across varying levels of LLM adaptation, from base models to fine-tuned variants. While the few-shot setting shows degradation in detection robustness, we attribute this to topic mismatch between the few-shot exemplars and the review being generated. This limitation can be mitigated with better exemplar selection or dynamic prompt construction.

From a deployment perspective, TBW offers a practical solution for reviewer attribution. The method is efficient and detection incurs minimal computational overhead, making it suitable for integration into existing conference submission pipelines~\cite{nemecek2024topic}. Its low latency and lack of architectural modifications make it a compelling candidate for enforcement mechanisms in venues that prohibit LLM-assisted review writing.

Lastly, our evaluation uses a constrained input (title and abstract) due to context window limitations. We expect that access to the full paper would further enhance generation quality and strengthen watermark consistency by grounding outputs in topic-relevant content.

\section{Peer Review Task Specifics}\label{appendixA}
This appendix provides additional details regarding the peer review generation setup described in \text{\S}\ref{genTask}. Specifically, we include conference-level review statistics and implementation details for TBW as well as prompting and fine-tuning the \texttt{Llama-3.1-8B} model.

\subsection{Conference Review Statistics}\label{appendA1}
Table~\ref{tab:review_stats} reports the number of reviews collected from each ICLR and NeurIPS conference used in our experiments. Only reviews submitted prior to the release of ChatGPT (November 2022) were included to minimize the likelihood of LLM-generated content in the training data. No additional filtering was applied beyond restricting the dataset to pre-ChatGPT conferences where all reviews were used in their original form.

\begin{table}[h!]
\centering
\begin{tabular}{|l|c|}
\hline
\textbf{Conference: Year} & \textbf{Number of Reviews} \\
\hline
ICLR: 2018 & 935 \\
ICLR: 2019 & 1419 \\
ICLR: 2020 & 2213 \\
ICLR: 2021 & 2594 \\
ICLR: 2022 & 2617 \\
ICLR: 2023 & 3793 \\
NeurIPS: 2021 & 2768 \\
NeurIPS: 2022 & 2824 \\
\hline
\end{tabular}
\caption{Review counts per conference used in training and evaluation. The total number of unique reviews is 19,163.}
\label{tab:review_stats}
\end{table}

\subsection{TBW Parameter Details}\label{watermarking_params_details}

\subsubsection{Implementation Components}
We use the same core components as the original TBW implementation to ensure consistency. Token and topic embeddings are computed using the \texttt{all-MiniLM-L6-v2} sentence embedding model~\cite{reimers-2020-multilingual-sentence-bert}. Topic extraction from input prompts is performed using \texttt{KeyBERT}~\cite{grootendorst2020keybert} for keyword-based topic identification.

\subsubsection{Topic Assignment \& Green List Construction}
Following the TBW framework, we partition the vocabulary into green lists based on semantic similarity to our predefined set of $K = 4$ domain-specific topics: \{\texttt{theory}, \texttt{applications}, \texttt{models}, \texttt{optimization}\}. Each token $v \in V$ is assigned to the green list $G_{t_i}$ if its cosine similarity with topic $t_i$ exceeds threshold $\tau$:
$$
\text{sim}(v, t_i) = \frac{\mathbf{e}_v \cdot \mathbf{e}_{t_i}}{\|\mathbf{e}_v\| \|\mathbf{e}_{t_i}\|}
$$

Tokens not meeting the similarity threshold are placed in a residual set and evenly distributed across all green lists to maintain complete vocabulary coverage.

\subsubsection{Topic Granularity Selection}
We selected $K = 4$ domain-specific topics to balance semantic coverage with watermark detection robustness. Finer-grained topic divisions (larger $K$) would create sparser green lists $G_{t_i}$, potentially weakening watermark signals and reducing robustness to paraphrasing. Coarser divisions (smaller $K$) would sacrifice semantic alignment between prompts and outputs. The selected topics: \{\texttt{theory}, \texttt{applications}, \texttt{models}, \texttt{optimization}\}, capture the major thematic dimensions of ML conference reviews while maintaining sufficient green-list density for reliable detection. Future work could explore adaptive topic granularity strategies, such as hierarchical topic structures or automatic topic discovery from review corpora.

\subsubsection{Generation \& Detection Parameters}
We apply a logit bias of $\delta = 2.0$ to green-list tokens during generation, consistent with values reported in prior watermarking literature~\cite{kirchenbauer2023watermark}. For token-to-topic assignment, we primarily use a cosine similarity threshold of $\tau = 0.7$, with additional evaluation at $\tau = 0.3$ to assess how watermark detection and text quality vary under relaxed alignment constraints.

For detection, we use the statistical z-test with baseline proportion $\gamma$ set to the expected green-token rate under random sampling, and threshold $z_{\text{threshold}}$ tuned for desired false positive rates.

\subsection{Model Configuration Details}\label{app:model_setup}
\subsubsection{Few-shot Prompting Setup} 
In the few-shot setting, the model is given a prompt containing a paper's title and abstract followed by a fixed instruction:
\begin{tcolorbox}[reviewbox]
\textbf{Title:} [TITLE] \\
\textbf{Abstract:} [ABSTRACT]\\
Please write a detailed review.
\end{tcolorbox}
\noindent Each prompt includes two example reviews prepended to help the model learn the expected structure and tone of a review. These few-shot examples are randomly sampled from the training pool but excluded from evaluation generations. Specifically, the two examples prepended to each prompt are drawn from the first two entries in the fine-tuning training split, ensuring consistency across models.

\subsubsection{Fine-Tuning Setup}\label{finetuning}
For fine-tuning, we follow a supervised instruction-tuning setup where each instance consists of an input prompt (title + abstract + instruction) and a target completion (review text). The dataset is split into training (80\%), validation (10\%), and test (10\%) subsets. We fine-tune using LoRA (Low-Rank Adaptation) with 4-bit quantization, enabling gradient checkpointing and early stopping. The objective is to improve the fluency and consistency of generated reviews while approximating the tone and structure typical of human-written peer reviews.

For instruction-tuned generation, we fine-tune the \texttt{Llama-3.1-8B} model using a parameter-efficient LoRA (Low-Rank Adaptation) method. LoRA freezes the original model weights and injects trainable low-rank matrices into a subset of layers, enabling effective fine-tuning with a small number of additional parameters. This approach is well-suited for large-scale models, reducing memory usage and training time while maintaining performance. Key settings are described in Table~\ref{tab:hyperparams}.
\begin{table}[h]
\centering
\small
\begin{tabular}{ll}
\toprule
\textbf{Parameter} & \textbf{Value} \\
\midrule
Adapter type & LoRA \\
LoRA $r$/$\alpha$ & 16/32 \\
LoRA dropout & 0.1 \\
Training epochs & 3 \\
Batch size (per device) & 2 \\
Max sequence length & 2048 tokens \\
Learning rate & 1e-4 \\
Warmup ratio & 0.2 \\
Quantization & 4-bit (NF4), double quantization \\
Target modules & \texttt{q\_proj}, \texttt{k\_proj}, \texttt{v\_proj}, \\
& \texttt{o\_proj}, \texttt{gate\_proj}, \texttt{up\_proj}, \\
& \texttt{down\_proj} \\
\bottomrule
\end{tabular}
\caption{Fine-tuning Hyperparameters}
\label{tab:hyperparams}
\end{table}
\noindent All experiments were conducted using the Hugging Face \texttt{Transformers} and \texttt{PEFT} libraries, with training orchestrated using the \texttt{Trainer} API. The final adapters and tokenizer were saved for downstream evaluation. The dataset consists of the prompt (title, abstract, and generation instruction) and a completion (review text), compatible with instruction tuning for causal language models.

\section{Generation Quality Evaluations}\label{appendixD}
We expand our evaluation of topic-based watermarking (TBW) to assess its sensitivity to different token-to-topic similarity thresholds. Specifically, we re-run perplexity and BERTScore evaluations using a lower semantic similarity threshold of $\tau = 0.3$ (vs.\ $\tau = 0.7$ in the main experiments). We also provide BERTScore comparisons of TBW ($\tau=0.7$) against KGW and SynthID baseline watermarking schemes.

\subsection{Baseline Watermarking Parameters}\label{baseline_params}
We compare TBW against two established watermarking methods: KGW~\cite{kirchenbauer2023watermark}, a pioneering approach for LLM watermarking, and SynthID-Text~\cite{dathathri2024scalable}, Google's proprietary technique for text attribution. All baseline implementations use the open-source MarkLLM framework~\cite{pan-etal-2024-markllm} with the following configurations:

\textbf{KGW Parameters:}
\begin{lstlisting}[style=paramstyle]
{
  "gamma": 0.5,
  "delta": 2.0,
  "prefix_length": 1,
  "z_threshold": 4.0
}
\end{lstlisting}
\textbf{SynthID Parameters:}
\begin{lstlisting}[style=paramstyle]
{
  "ngram_len": 5,
  "watermark_mode": non-distortionary,
  "threshold": 0.52,
  "detector_type": mean
}
\end{lstlisting}



\subsection{BERTScore Evaluation (TBW $\tau = 0.7$)}\label{bertbert}
\begin{figure}[h!]
\begin{center}
\centerline{\includegraphics[width=1\columnwidth]{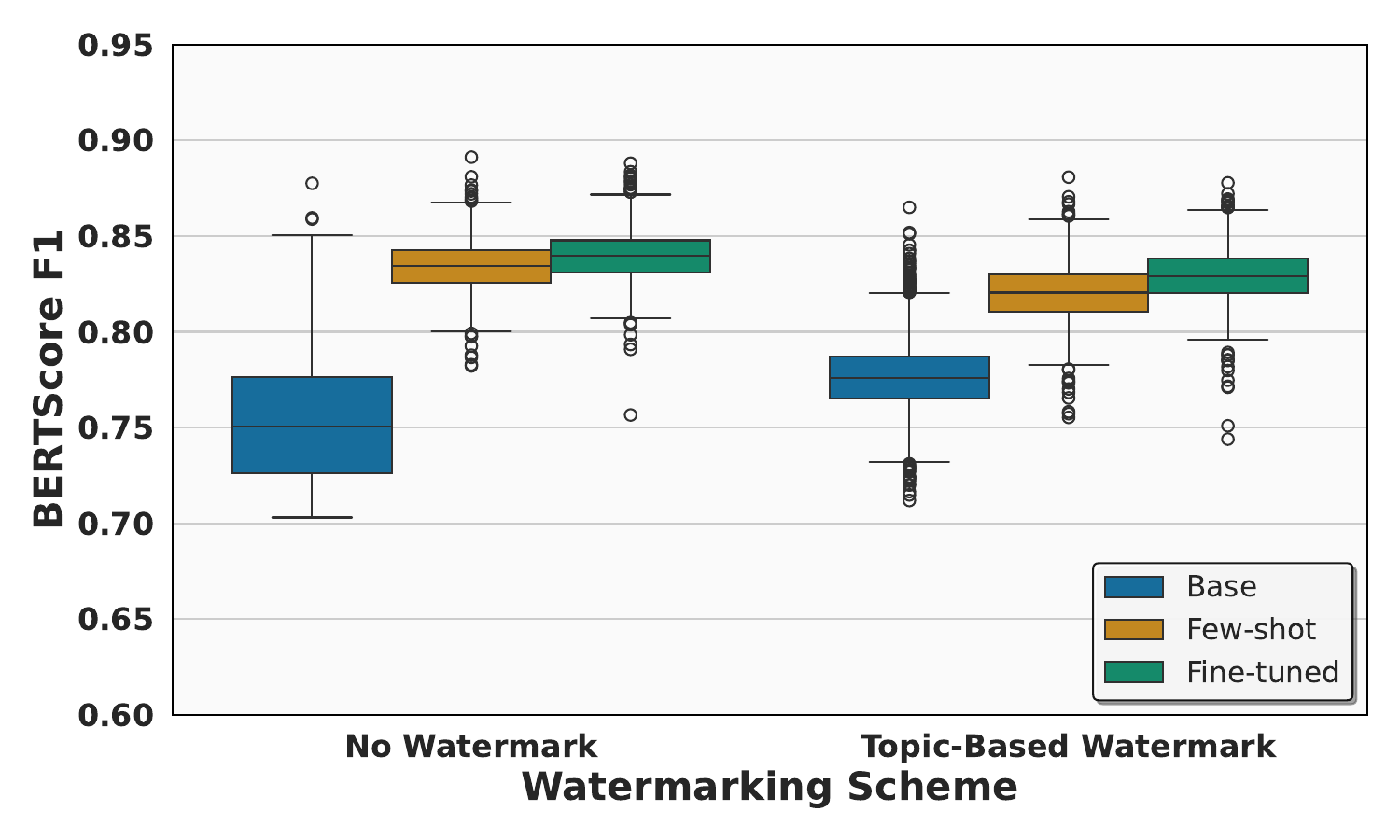}}
\caption{BERTScore F1 distributions across model configurations with and without TBW ($\tau = 0.7$). Higher values indicate greater semantic similarity to the ground truth.}
\label{bert-07}
\end{center}
\end{figure}

We use BERTScore F1 to evaluate semantic similarity between generated reviews and ground-truth references. This metric compares contextual embeddings and is tolerant to paraphrasing, making it well-suited for open-ended review generation. Results across all model configurations are shown in Figure~\ref{bert-07}.

TBW causes only a minor drop in BERTScore, indicating that semantic fidelity is largely preserved. Notably, in the base model, TBW narrows the BERTScore distribution, suggesting more consistent alignment with the source prompt across samples.

\subsection{Human Evaluation}\label{humanevals}
To complement our automated text quality metrics, we conduct a small-scale human evaluation to assess the perceptual quality of TBW-generated peer reviews. Human evaluation provides crucial insights into whether watermark artifacts are detectable to end users, which is essential for practical deployment.

We conduct a single-set evaluation with 6 human evaluators to assess individual text samples across multiple quality dimensions. All evaluators came from a computer science background and were familiar with the review process of conferences used in our few-shot and fine-tuned model training (i.e., ICLR and NeurIPS). 

Evaluators were presented with 15 generated review samples under our standard configuration ($\tau = 0.7$, \texttt{Llama-3.1-8B}). For each sample, evaluators were shown the paper title, abstract, and generated review text. Of the 15 samples, 5 were generated using the base model, 5 using few-shot, and 5 using fine-tuned configurations. Sample order was randomized to avoid ordering effects. Evaluators rated each review on a 5-point Likert scale (1 = very poor, 5 = excellent) across three dimensions:
\begin{itemize}
    \item \textbf{Fluency}: ``How fluent and natural is the text?''
    \item \textbf{Coherence}: ``How logically consistent and easy to follow is this text?''
    \item \textbf{Usefulness}: ``How useful would this review be for authors and reviewers?''
\end{itemize}

Table~\ref{tab:human_eval} summarizes the human evaluation results across all quality dimensions. As hypothesized, the base model performed worst across all metrics, with few-shot showing intermediate performance and fine-tuned achieving the highest scores. The fine-tuned model achieved mean ratings of 3.87 for fluency, 3.80 for coherence, and 3.60 for usefulness, demonstrating that domain adaptation via fine-tuning produces reviews perceived as substantially higher quality by human evaluators.

\begin{table}[h!]
\centering
\small
\begin{tabular}{lccc}
\toprule
\textbf{Model} & \textbf{Fluency} & \textbf{Coherence} & \textbf{Usefulness} \\
\midrule
Base       & 2.80 $\pm$ 1.16 & 2.47 $\pm$ 0.97 & 2.07 $\pm$ 1.05 \\
Few-shot   & 3.70 $\pm$ 0.65 & 3.47 $\pm$ 0.73 & 2.87 $\pm$ 1.07 \\
Fine-tuned & \textbf{3.87 $\pm$ 0.78} & \textbf{3.80 $\pm$ 0.76} & \textbf{3.60 $\pm$ 0.97} \\
\bottomrule
\end{tabular}
\caption{Human evaluation results for TBW-generated reviews across model configurations. Scores are on a 5-point Likert scale (mean $\pm$ std. dev., $n=6$ evaluators, 5 samples per configuration).}
\label{tab:human_eval}
\end{table}

One notable result is the usefulness rating of few-shot samples. While few-shot generation shows substantial improvements in fluency (3.70) and coherence (3.47) compared to the base model, usefulness remains relatively lower at 2.87. This pattern aligns with our findings in Section~\ref{robust}, where few-shot models exhibited degraded watermark detection performance under paraphrasing attacks. We hypothesize that topic misalignment between prompt exemplars and target papers weakens both topical consistency (affecting detection robustness) and review relevance (affecting perceived usefulness). Despite this limitation, the few-shot approach still produces reviews that human evaluators rate as reasonably fluent and coherent, suggesting that the watermarking mechanism itself does not introduce perceptible artifacts that degrade text quality.

\subsection{BERTScore Evaluation (KGW \& SynthID)}\label{lowerbertbert}
\begin{figure}[h!]
\begin{center}
\centerline{\includegraphics[width=1\columnwidth]{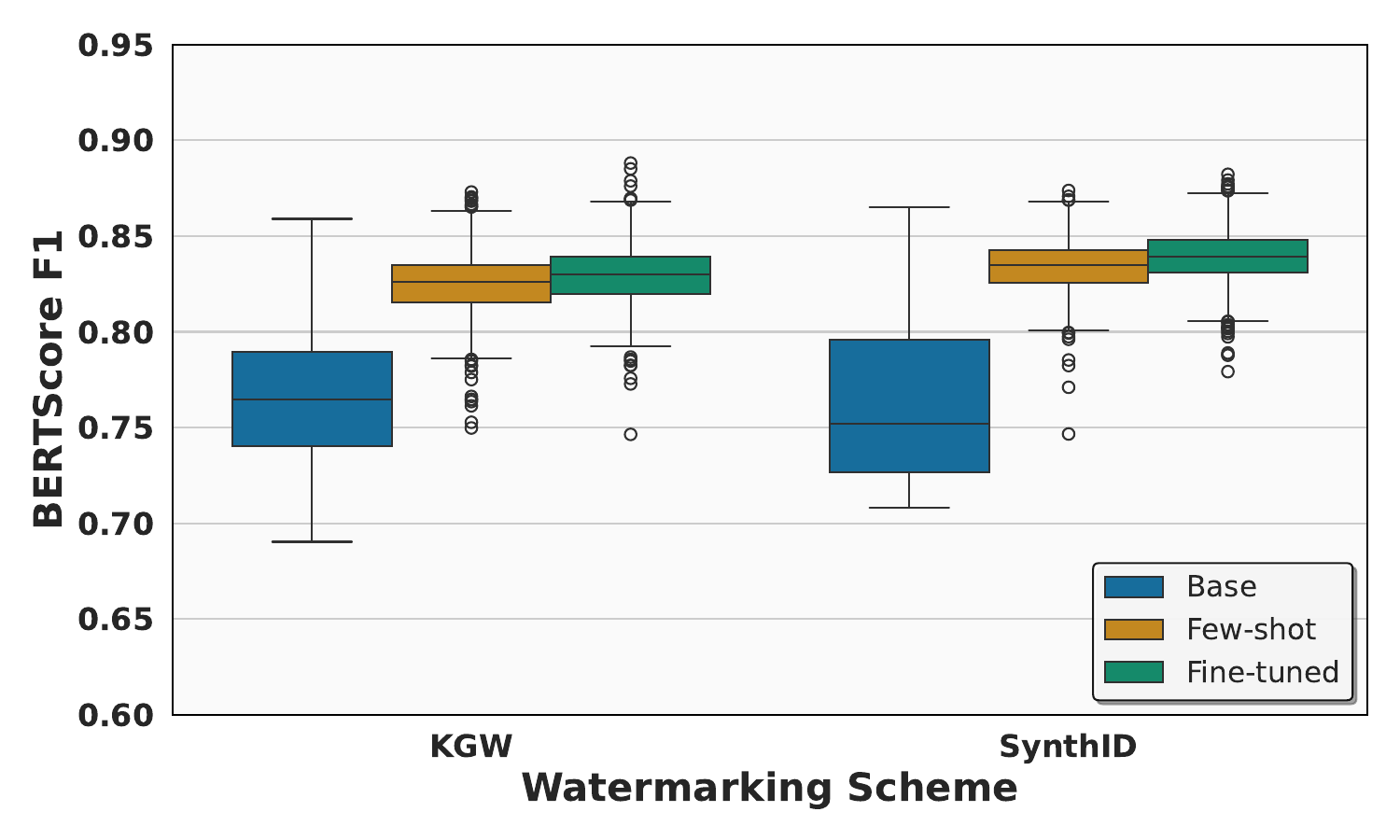}}
\caption{
BERTScore F1 distributions across model configurations with KGW and SynthID. Higher values indicate greater semantic similarity to the human-written reference.
}
\label{bert-other}
\end{center}
\end{figure}

We evaluate BERTScore F1 for generations produced with KGW and SynthID. Results are presented in Figure~\ref{bert-other}.

In the few-shot and fine-tuned configurations, KGW performs comparably to TBW at $\tau = 0.7$, with similar median BERTScore values and distributional tightness. However, in the base model configuration, KGW shows a broader distribution of scores, indicating higher variability in semantic fidelity. This suggests that KGW, like TBW, is more effective when the generation is guided by conditioning or domain adaptation. SynthID shows a similar pattern but with slightly more pronounced effects. In the base model, SynthID outputs exhibit a wider spread compared to both TBW and KGW, reflecting less stable semantic alignment. In contrast, SynthID performs slightly better in the few-shot and fine-tuned settings, with a 1-2\% improvement in BERTScore F1 over TBW at $\tau = 0.7$.

These results highlight that while all watermarking methods introduce some tradeoff between attribution and quality, their semantic fidelity is more stable in strongly conditioned generation settings. SynthID offers stronger semantic preservation under tight generation constraints, but at the cost of higher perplexity and fluency degradation in lower-context scenarios.

\subsection{Evaluation with Lower Topic Similarity Threshold ($\tau=0.3$)}\label{lowerQuality}
We repeat the perplexity and BERTScore evaluations using a relaxed topic assignment threshold of $\tau = 0.3$. This setting allows more tokens to be included in each green list, resulting in stronger watermark signals but potentially greater degradation in generation quality. The results help assess how sensitive TBW is to this design parameter.

\subsubsection{Perplexity}
\begin{figure}[h!]
\begin{center}
\centerline{\includegraphics[width=1\columnwidth]{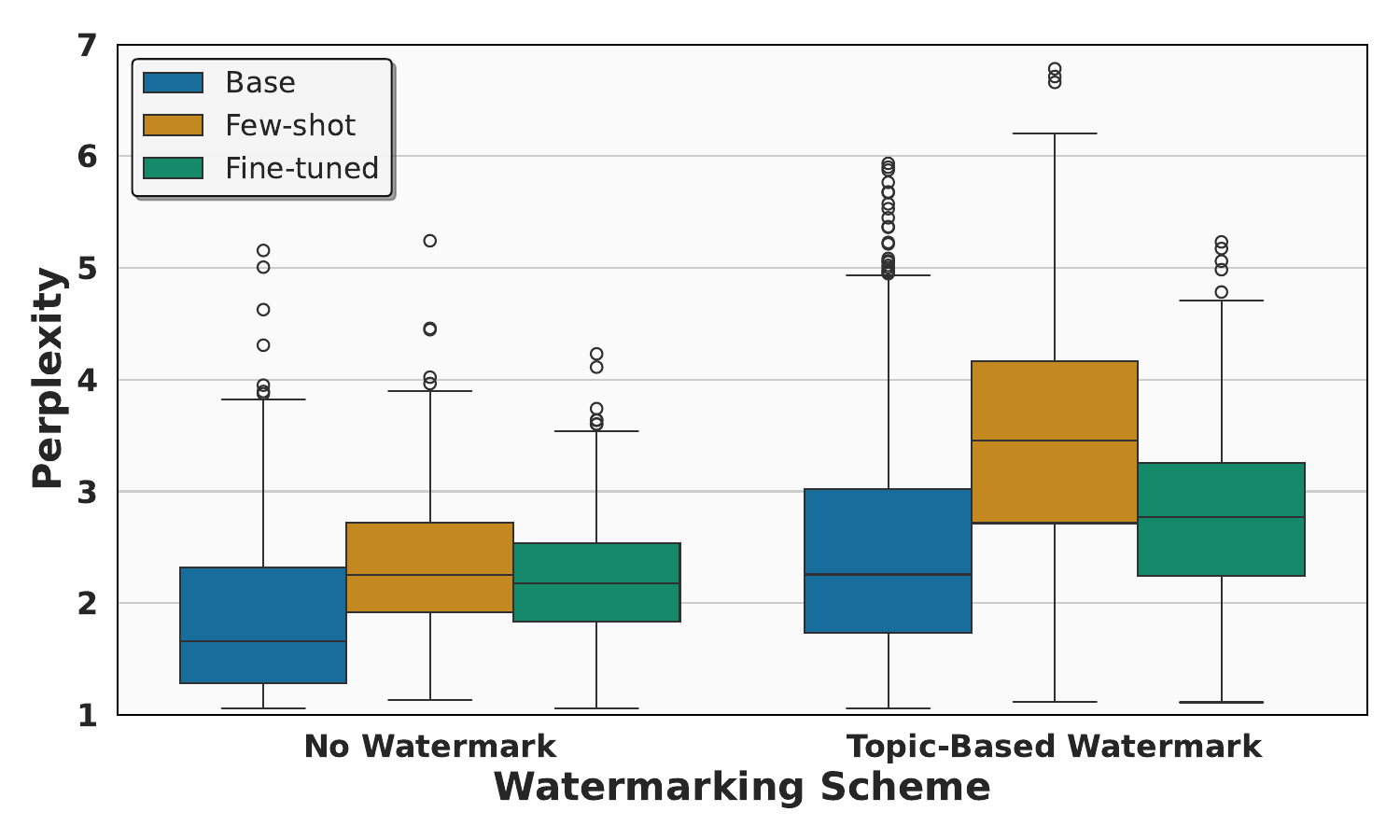}}
\caption{Perplexity distributions across model configurations with and without TBW ($\tau = 0.3$). Lower values indicate better fluency. Values above 20 are truncated for clarity.
}
\label{perplexity-03}
\end{center}
\end{figure}

Figure~\ref{perplexity-03} shows the perplexity distributions for all model configurations, comparing outputs generated with and without TBW under $\tau = 0.3$. Following the same visualization protocol as in the main paper, we truncate values above 20 for readability. Table~\ref{holdout03} reports how many samples remained below this threshold in each setting.

\begin{table}[h!]
\centering
\begin{tabular}{llc}
\toprule
\textbf{Model} & \textbf{Scheme} & \textbf{Samples Retained} \\
\midrule
\multirow{2}{*}{Base}      & NW & 508  \\
                           & TBW & \textbf{684} \\
                           \cmidrule(){1-3}
\multirow{2}{*}{Few-shot}   & NW & \textbf{1000}  \\
                           & TBW & \textbf{1000} \\
                           \cmidrule(){1-3}
\multirow{2}{*}{Fine-tuned} & NW & \textbf{1000}  \\
                           & TBW & \textbf{1000} \\
\bottomrule
\end{tabular}
\caption{Number of generations with perplexity $\leq 20$, comparing unwatermarked (NW) and TBW outputs ($\tau = 0.3$).
}
\label{holdout03}
\end{table}
As expected, TBW at $\tau = 0.3$ produces slightly higher perplexity than unwatermarked generations, reflecting modest fluency degradation. Compared to TBW at $\tau = 0.7$, this lower-threshold variant results in fewer retained samples in the base model (684 vs. 991), suggesting increased fluency loss under weaker semantic filtering. Additionally, there is worse performance in the few-shot model, consistent with less effective topic alignment, but with improved perplexity in the fine-tuned model potentially due to the broader green lists better overlap with the model's learned domain-specific vocabulary.

These results support the view that $\tau$ serves as a tradeoff between watermark strength and generation quality, and that optimal settings may vary depending on the model's adaptation level.

\subsubsection{BERTScore Evaluation}
\begin{figure}[h!]
\begin{center}
\centerline{\includegraphics[width=1\columnwidth]{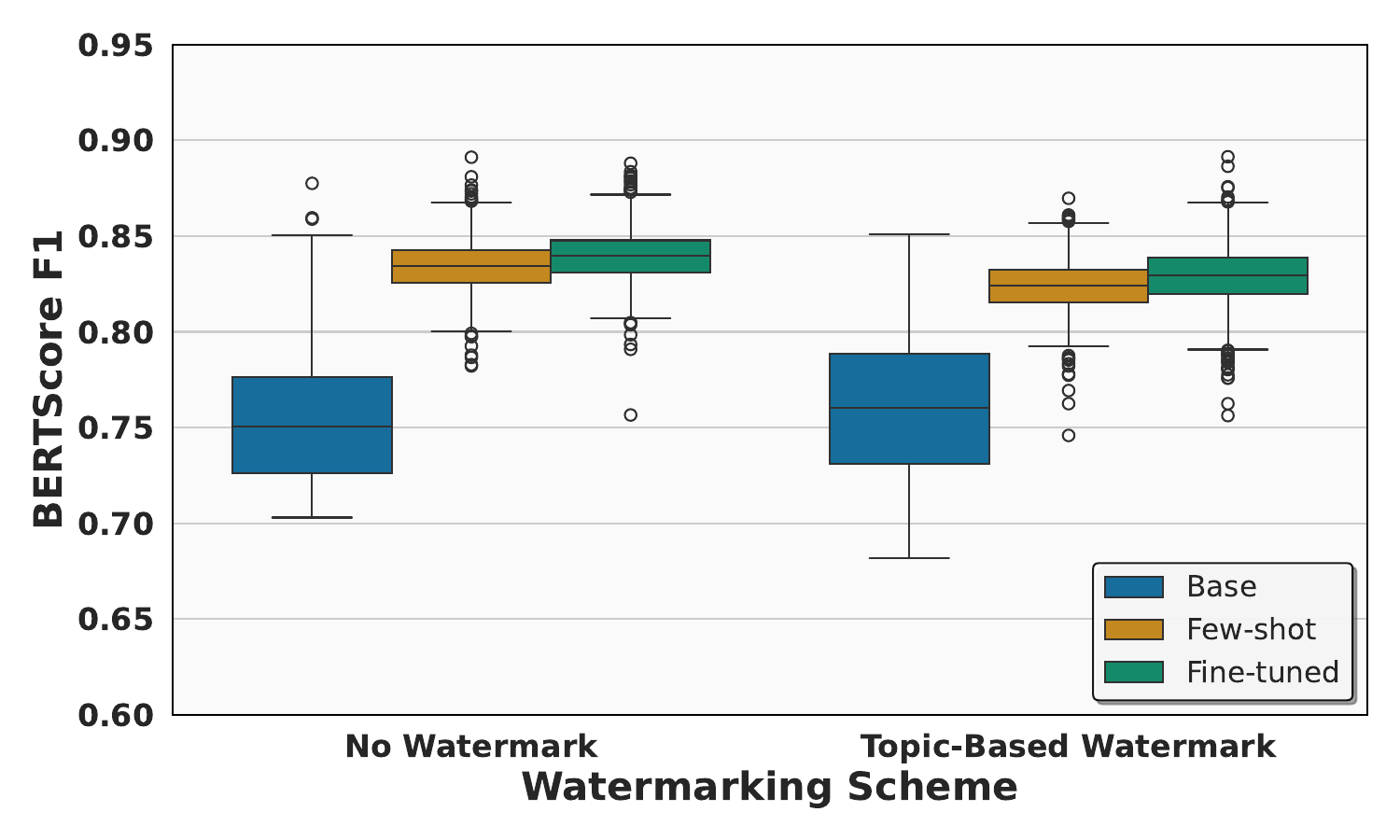}}
\caption{
BERTScore F1 distributions across model configurations with and without TBW ($\tau = 0.3$). Higher values indicate greater semantic similarity to the human-written reference.
}
\label{bert-03}
\end{center}
\end{figure}

We repeat the BERTScore F1 evaluation, using generations produced with TBW at $\tau = 0.3$. Results are shown in Figure~\ref{bert-03}.

We observe that TBW with $\tau = 0.3$ results in similar BERTScore degradation as seen with $\tau = 0.7$ in both the few-shot and fine-tuned model configurations. This indicates that semantic fidelity is largely preserved even with a broader green list, suggesting the robustness of TBW's semantic biasing strategy in these more guided generation settings.

However, the base model configuration shows more pronounced differences. Compared to TBW at $\tau = 0.7$, the base model with $\tau = 0.3$ produces generations with a broader range of BERTScore values, indicating increased variability in semantic alignment. This dispersion suggests that, in the absence of stronger conditioning (e.g., few-shot or fine-tuning), relaxing the similarity threshold introduces more topical drift, potentially reducing TBW's ability to maintain consistent semantic guidance.

These results reinforce that TBW is more stable in controlled generation setups, while its performance in lower-context settings (like the base model) is more sensitive to the choice of $\tau$.

\section{Robustness Evaluations}\label{appendE}
We provide comprehensive robustness analysis for the evaluations described in \text{\S}\ref{robust}, including detailed performance metrics and ROC curve analysis across all experimental conditions.

\begin{figure}[h!]
\begin{center}
\centerline{\includegraphics[width=1\columnwidth]{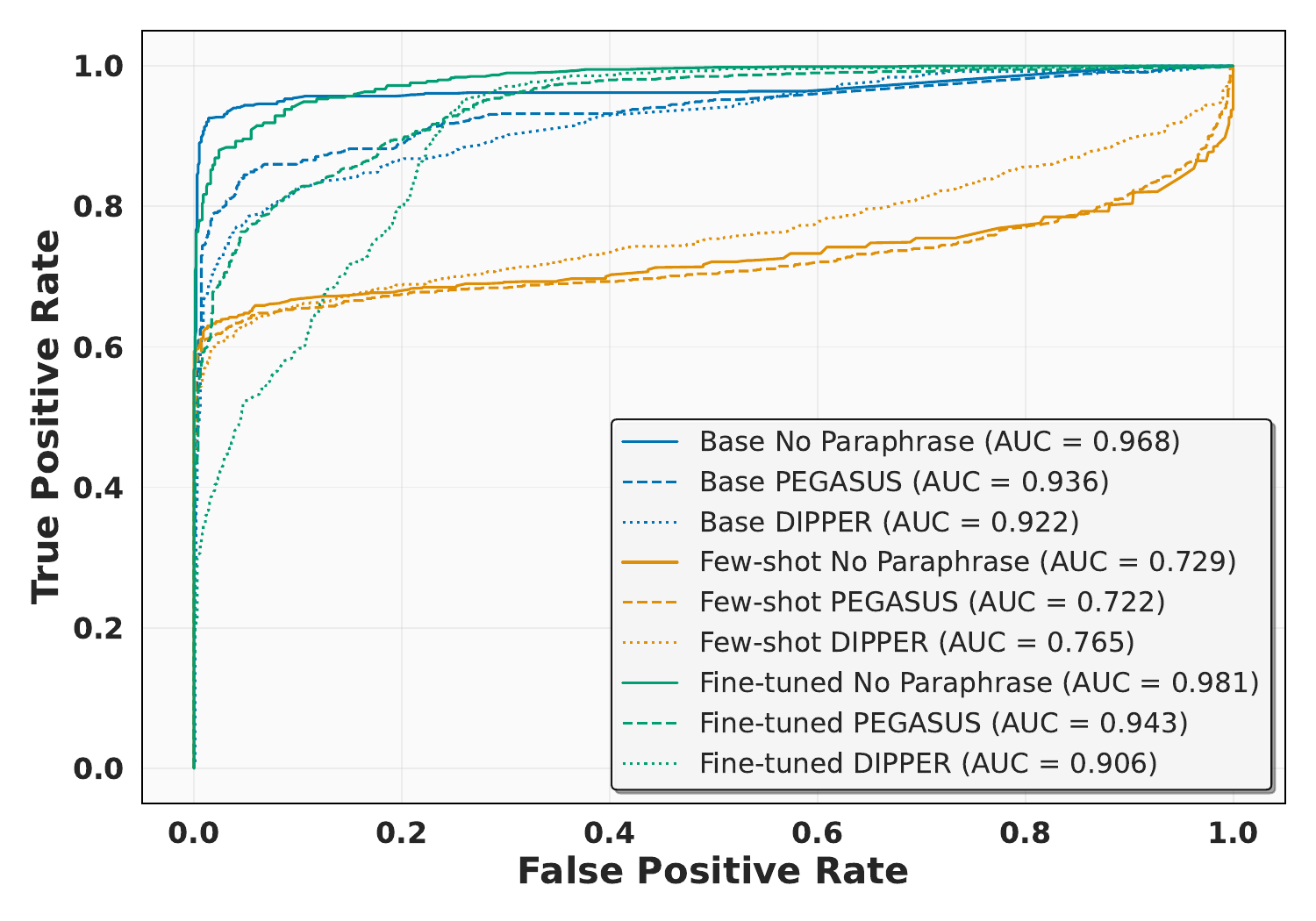}}
\caption{ROC curves for TBW detection under no attack, PEGASUS, and DIPPER paraphrasing, across all model configurations. The curves demonstrate TBW's robustness across attack severity and adaptation settings.}
\label{roc_curve}
\end{center}
\end{figure}

Figure~\ref{roc_curve} presents ROC curves for TBW detection performance across base, few-shot, and fine-tuned model configurations under three conditions: no attack baseline, PEGASUS paraphrasing, and DIPPER paraphrasing. The curves demonstrate TBW's consistent robustness across varying attack severity and model adaptation settings. Detection performance remains robust in base and fine-tuned configurations, with area under the curve (AUC) values exceeding 0.90 under no attack conditions and experiencing only moderate degradation under paraphrasing attacks. The few-shot model exhibits greater sensitivity to paraphrasing-induced topic dilution, as discussed in \text{\S}\ref{robust}, resulting in reduced detection confidence and lower overall performance across all attack conditions.

\begin{table}[ht]
\centering
\scriptsize
\begin{tabular}{@{}llcccc@{}}
\toprule
\textbf{Model} & \textbf{Attack} & \textbf{AUC} & \textbf{F1} & \textbf{TPR@1\%} & \textbf{TPR@10\%} \\
\hline
\multirow{3}{*}{Base} 
    & None     & 0.968 & 0.955 & 0.908 & 0.956 \\
    & PEGASUS  & 0.936 & 0.893 & 0.746 & 0.861 \\
    & DIPPER   & 0.922 & 0.857 & 0.669 & 0.826 \\
\hline
\multirow{3}{*}{Few-shot} 
    & None     & 0.729 & 0.768 & 0.626 & 0.669 \\
    & PEGASUS  & 0.722 & 0.758 & 0.609 & 0.655 \\
    & DIPPER   & 0.765 & 0.754 & 0.565 & 0.659 \\
\hline
\multirow{3}{*}{Fine-tuned} 
    & None     & 0.981 & 0.927 & 0.817 & 0.948 \\
    & PEGASUS  & 0.944 & 0.858 & 0.593 & 0.826 \\
    & DIPPER   & 0.906 & 0.861 & 0.348 & 0.598 \\
\bottomrule
\end{tabular}
\caption{Detection performance across model configurations and attack settings. Metrics include ROC-AUC, best F1 score, and true positive rate (TPR) at fixed false positive rates (FPRs) of 1\% and 10\%.}
\label{evaluation_metrics}
\end{table}

Table~\ref{evaluation_metrics} provides comprehensive detection performance across all experimental conditions, including ROC-AUC, F1 scores, and true positive rates (TPR) at relevant false positive rate (FPR) thresholds of 1\% and 10\%. The base model demonstrates robustness, maintaining an AUC above 0.92 even under the strongest paraphrasing attacks, with F1 scores of 0.893 (PEGASUS) and 0.857 (DIPPER). At the 1\% FPR threshold, the base model achieves TPR values of 74.6\% and 66.9\% under PEGASUS and DIPPER attacks, respectively, indicating strong practical utility for high-precision detection scenarios.

The fine-tuned model shows similar resilience with AUC values of 0.944 (PEGASUS) and 0.906 (DIPPER), though it exhibits greater sensitivity at low FPR thresholds, particularly under DIPPER attacks where TPR@1\% drops to 34.8\%. This suggests that fine-tuning may increase vulnerability to certain paraphrasing patterns while maintaining overall detection capability.

In contrast, the few-shot configuration shows limited degradation under attacks, with AUC values remaining stable around 0.72-0.77 across all conditions. However, the consistently lower baseline performance (AUC = 0.729) indicates that topic misalignment limits detectability in this setting, making the relative robustness less operationally significant.

The maintained performance at low FPR thresholds across model configurations confirms that TBW's vocabulary partitioning strategy effectively preserves detection capability while minimizing false alarms on human-written content, as evidenced by the consistent TPR@1\% and TPR@10\% metrics across experimental conditions.

\section{Classifier Specifics}\label{classifier-training}
We provide implementation details for the classification experiments described in \text{\S}\ref{class-based}. We outline the training setup used for both BERT and RoBERTa classifiers and summarize the evaluation strategy for attribution analysis on generated peer reviews.

\subsection{Data Construction \& Protocol}
We first construct a labeled dataset by extracting review texts from our generation pipeline and assigning a class label based on the associated ground truth rating (e.g., scores 1-4 mapped to \texttt{reject}, 5-6 to \texttt{borderline}, and 7-10 to \texttt{accept}). To ensure accurate mapping, we align generated reviews with their original metadata using paper titles as unique identifiers. The final dataset consists of generated reviews paired with class labels, drawn from the fine-tuned generation split described in \text{\S}\ref{finetuning}.

The final dataset consists of generated reviews paired with class labels, drawn from our experimental pipeline. The dataset is stratified into training and held-out test splits, with 9,000 balanced training samples (3,000 per class) and 1,000 test samples for evaluation.

\subsection{Classifier Training}
For reproducibility, we provide the specific training parameters used to fine-tune our LLM classifiers for predicting peer review labels corresponding to paper rating categories: \texttt{reject}, \texttt{borderline}, and \texttt{accept}.

Each model is fine-tuned using the Hugging Face \texttt{Trainer} API with early stopping based on F1. Key training settings described in Table~\ref{tab:classification-hyperparams}.
\begin{table}[h]
\centering
\small
\begin{tabular}{ll}
\toprule
\textbf{Parameter} & \textbf{Value} \\
\midrule
Model types & \texttt{bert-base-uncased}, \\ & \texttt{roberta-large} \\
Number of classes & 3 (\texttt{reject}, \texttt{borderline}, \texttt{accept}) \\
Max sequence length & 512 tokens \\
Training epochs & 5 \\
Batch size (per device) & 16 \\
Learning rate & 2e-5 \\
Warmup ratio & 0.1 \\
Optimizer & AdamW \\
Scheduler & Cosine with restarts \\
Dropout & 0.2 (attention and hidden layers) \\
Gradient clipping & Max norm 1.0 \\
Label smoothing & 0.1 \\
Precision & Mixed (FP16 with full-eval) \\
Quantization & 4-bit weight loading \\
Evaluation strategy & Per epoch; \\ & best model selected via F1 \\
Early stopping & Enabled (patience = 1) \\
\bottomrule
\end{tabular}
\caption{Classifier Training Hyperparameters}
\label{tab:classification-hyperparams}
\end{table}
\noindent Tokenization was performed using each model's pretrained tokenizer. A padding-aware data collator was used for batch construction. All training was conducted using the Hugging Face \texttt{Transformers} library and saved checkpoints were used for downstream evaluation on generated samples.

\subsection{Classifier Evaluation}
We evaluate both BERT and RoBERTa classifiers on a held-out test set of 1,000 human-written peer reviews. This evaluation step assesses whether the models can correctly recover the original review rating category (\texttt{reject}, \texttt{borderline}, \texttt{accept}) before applying them to generated or watermarked samples.

\begin{figure}[h!]
\begin{center}
\centerline{\includegraphics[width=1\columnwidth]{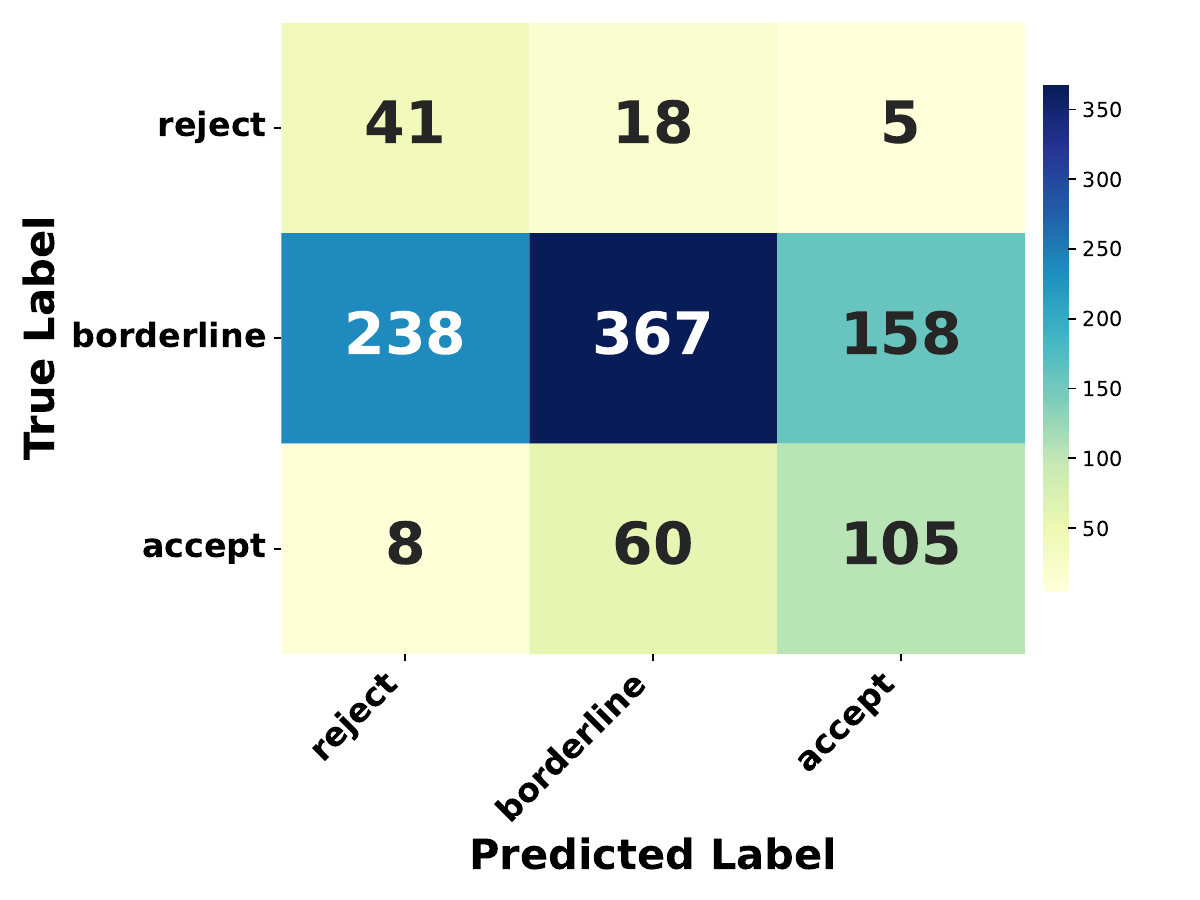}}
\caption{Confusion matrix for the BERT classifier on 1,000 human-written peer reviews.}
\label{bertclass}
\end{center}
\end{figure}

\begin{figure}[h!]
\begin{center}
\centerline{\includegraphics[width=1\columnwidth]{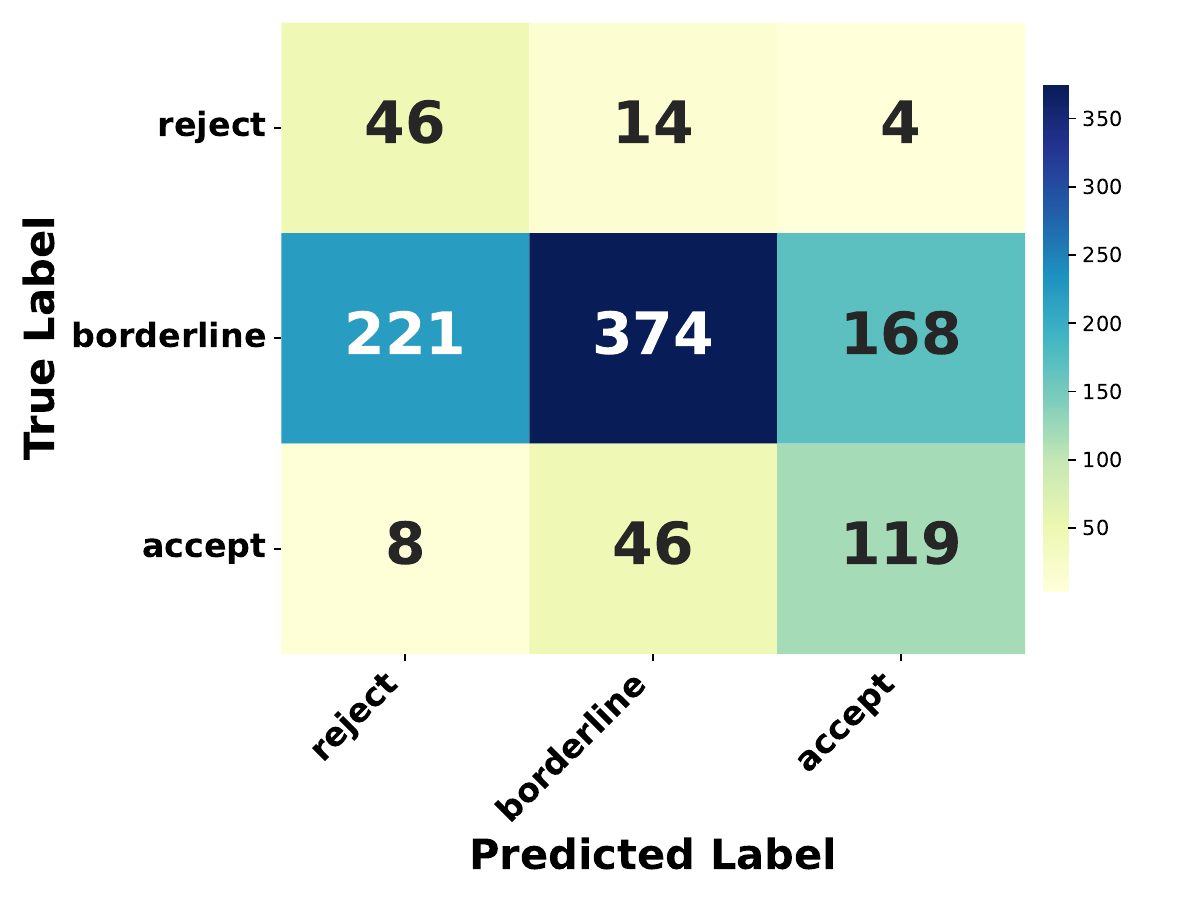}}
\caption{Confusion matrix for the RoBERTa classifier on 1,000 human-written peer reviews.}
\label{robertclass}
\end{center}
\end{figure}

Predictions are obtained from each trained classifier on the tokenized test set and compared against the ground truth labels. We compute confusion matrices to visualize class-specific misclassification patterns and report overall accuracy as a coarse measure of performance. BERT achieves an accuracy of 51.3\%, while RoBERTa performs slightly better at 53.9\%. Figures~\ref{bertclass} and~\ref{robertclass} present the confusion matrices for BERT and RoBERTa, respectively.

Both classifiers exhibit a strong predictive tendency toward the \texttt{borderline} class. As shown in the confusion matrices, the majority of \texttt{borderline} samples are correctly classified by both BERT (367/763) and RoBERTa (374/763). However, a large number of \texttt{reject} and \texttt{accept} samples are also misclassified as \texttt{borderline}. For instance, BERT misclassifies 18 \texttt{reject} and 60 \texttt{accept} samples as \texttt{borderline}, while RoBERTa reduces this to 14 and 46, respectively. Compared to BERT, RoBERTa shows slightly improved separation between all three classes, with fewer misclassifications across off-diagonal entries. In particular, it shows higher retention of true \texttt{reject} and \texttt{accept} labels, suggesting better overall discriminative performance.

\subsection{Class-Specific Classifier Evaluation}\label{classifieraddition}
To further characterize classifier performance, we conduct a comprehensive class-specific evaluation of generated peer reviews based on the classification framework introduced in \text{\S}\ref{class-based}. This analysis extends the aggregate metrics reported in Table~\ref{tab:overall-performance} by examining model behavior across the three target rating categories under different watermarking conditions and topic similarity thresholds.

\begin{figure*}[htbp]
    \centering
    \resizebox{0.8\textwidth}{!}{ 
    \begin{minipage}{\textwidth}
    \begin{subfigure}[b]{0.32\textwidth}
        \includegraphics[width=\textwidth]{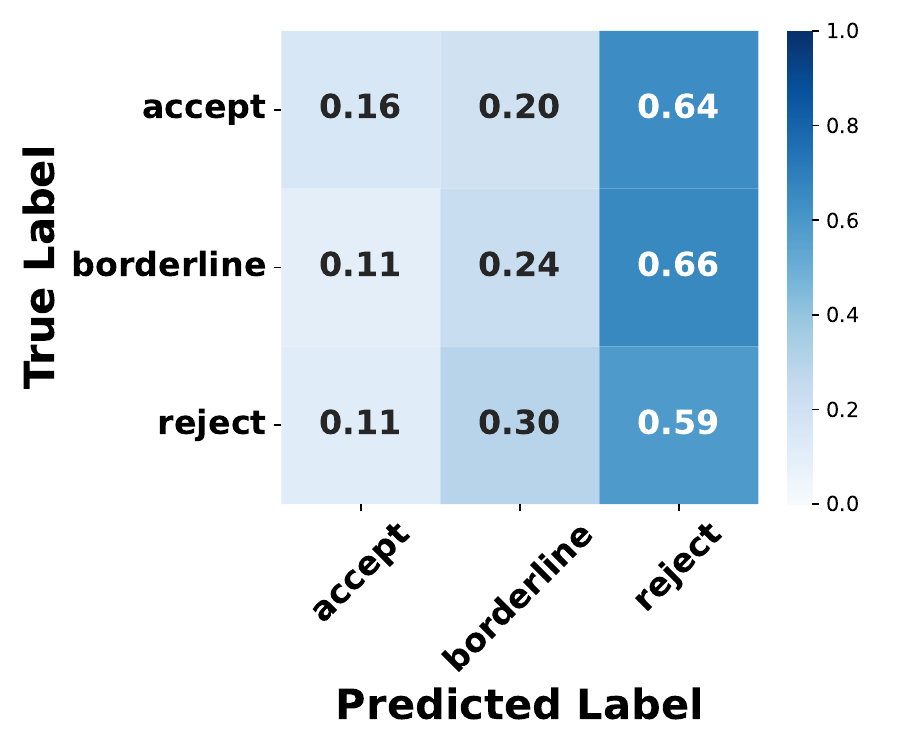}
        \caption{BERT Base NW}
    \end{subfigure}
    \begin{subfigure}[b]{0.32\textwidth}
        \includegraphics[width=\textwidth]{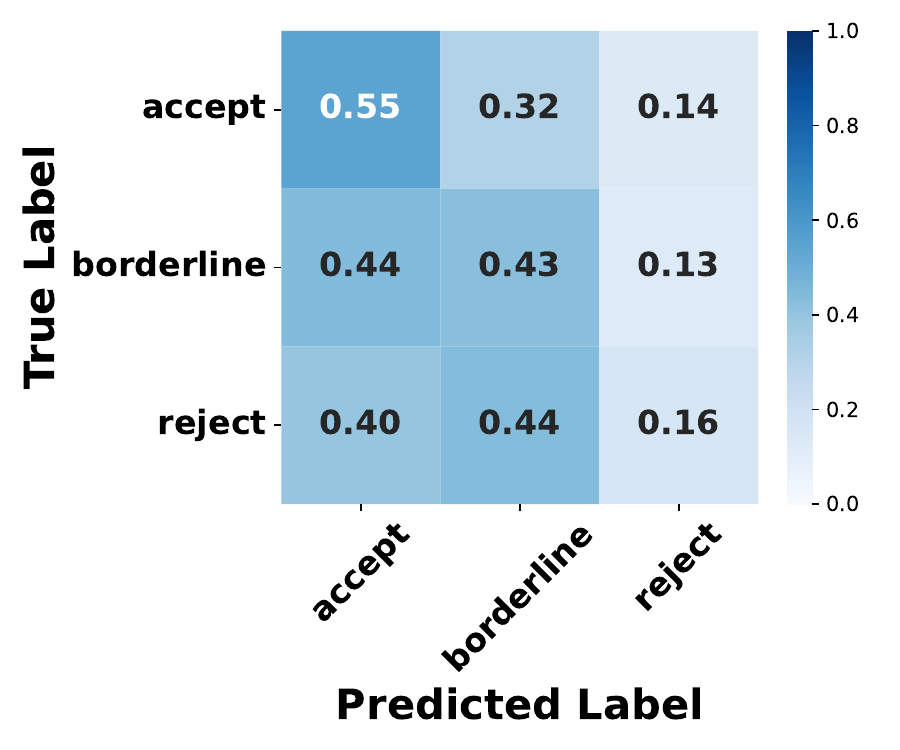}
        \caption{BERT Few-shot NW}
    \end{subfigure}
    \begin{subfigure}[b]{0.32\textwidth}
        \includegraphics[width=\textwidth]{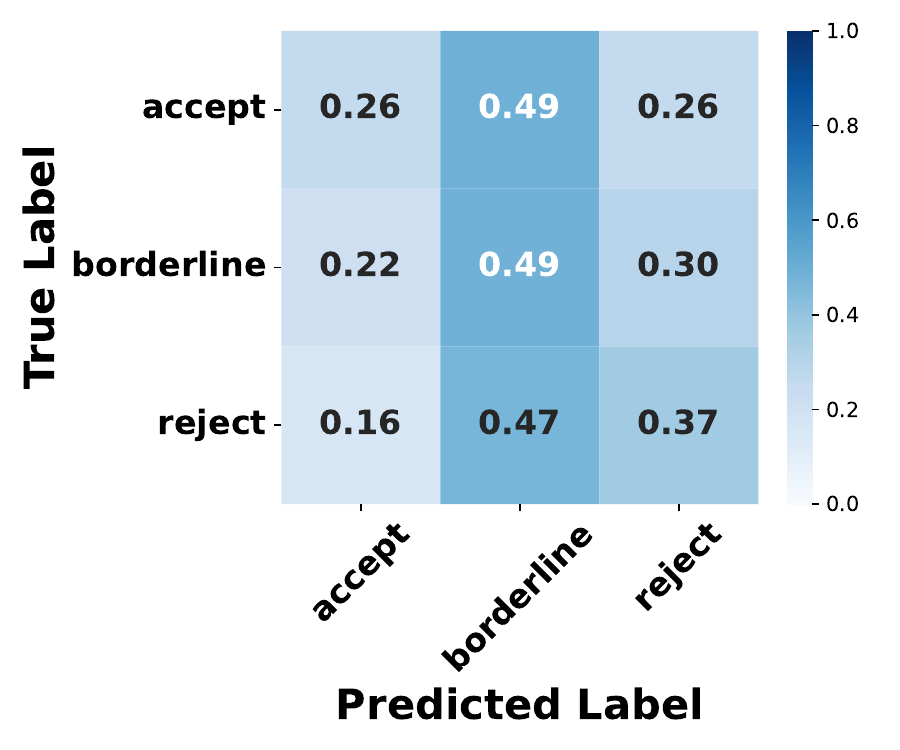}
        \caption{BERT Fine-tuned NW}
    \end{subfigure}
    
    \begin{subfigure}[b]{0.32\textwidth}
        \includegraphics[width=\textwidth]{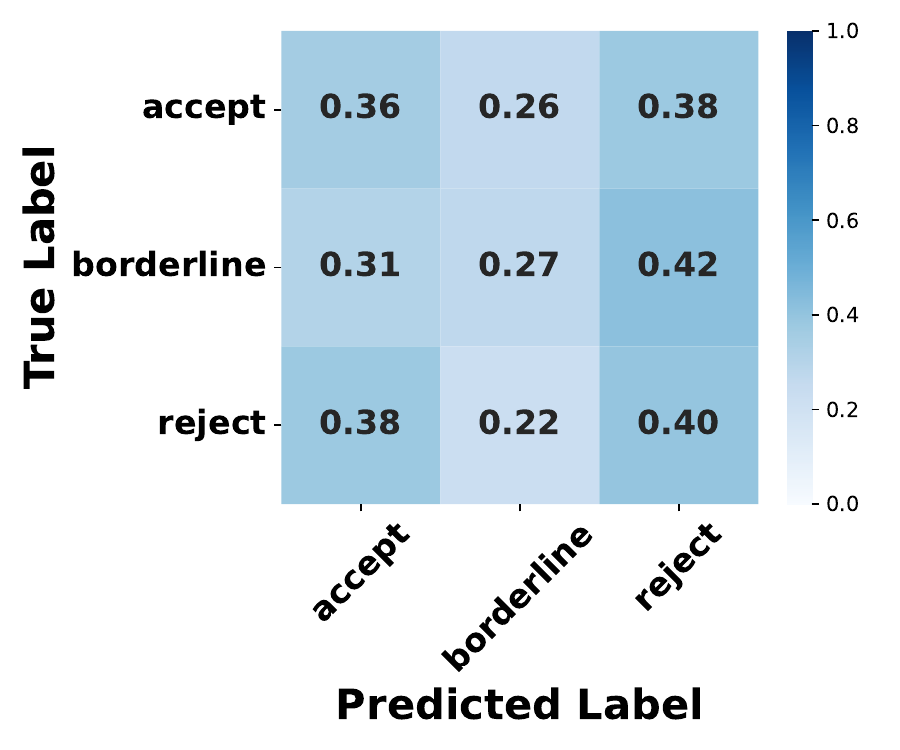}
        \caption{BERT Base TBW ($\tau=0.7$)}
    \end{subfigure}
    \begin{subfigure}[b]{0.32\textwidth}
        \includegraphics[width=\textwidth]{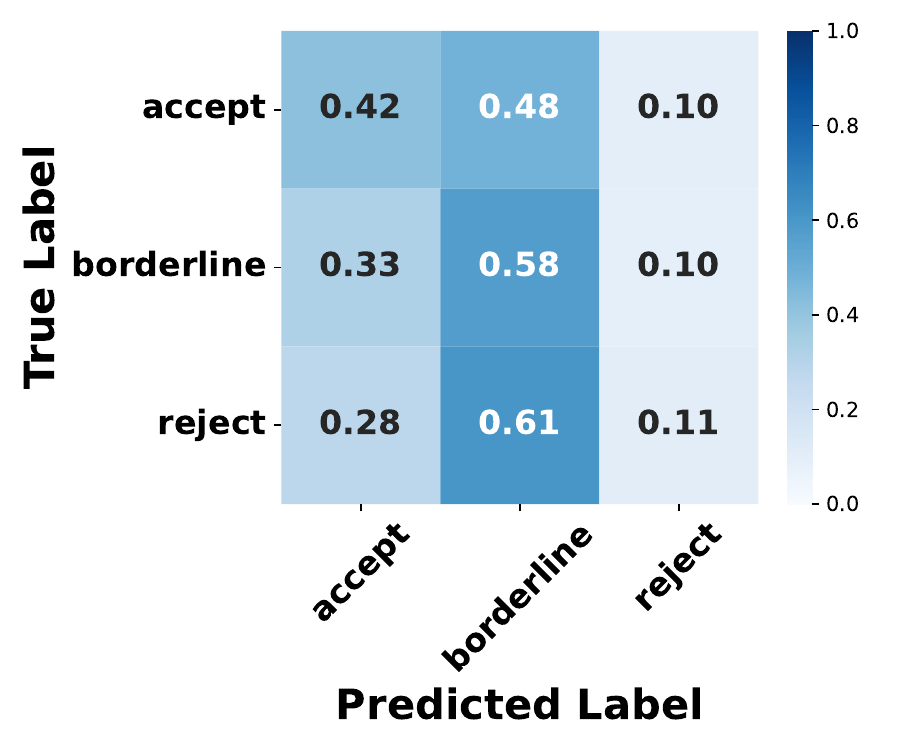}
        \caption{BERT Few-shot TBW ($\tau=0.7$)}
    \end{subfigure}
    \begin{subfigure}[b]{0.32\textwidth}
        \includegraphics[width=\textwidth]{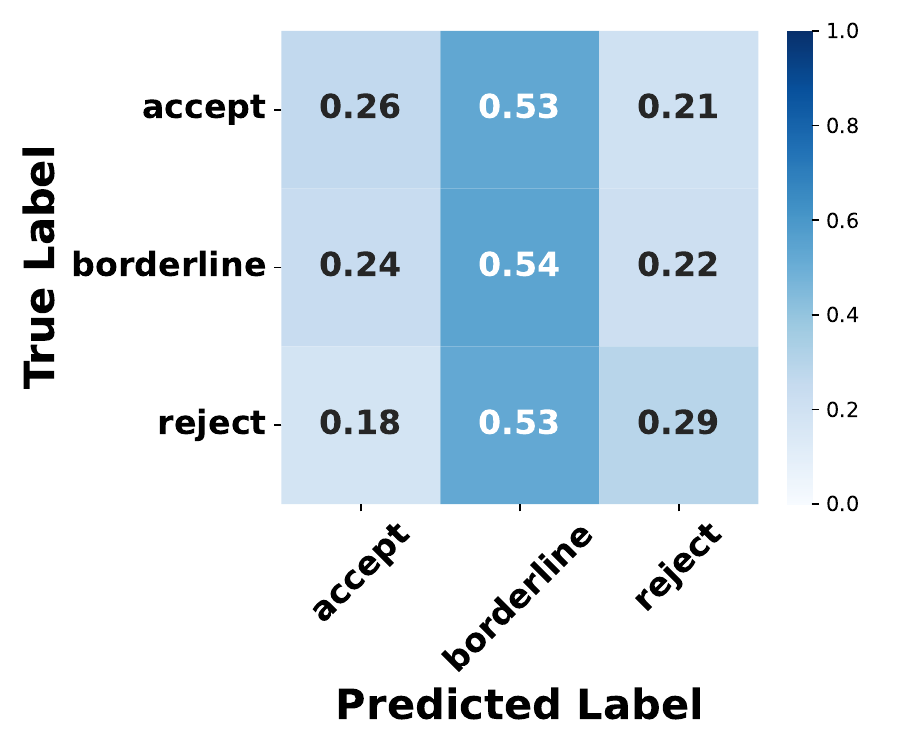}
        \caption{BERT Fine-tuned TBW ($\tau=0.7$)}
    \end{subfigure}
    
    \begin{subfigure}[b]{0.32\textwidth}
        \includegraphics[width=\textwidth]{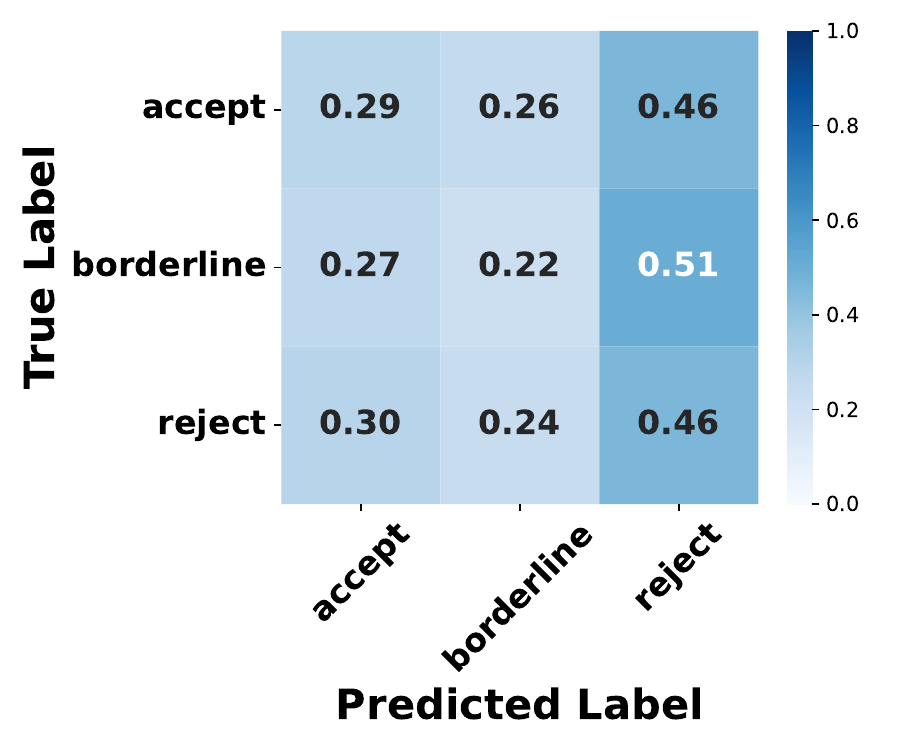}
        \caption{BERT Base TBW ($\tau=0.3$)}
    \end{subfigure}
    \begin{subfigure}[b]{0.32\textwidth}
        \includegraphics[width=\textwidth]{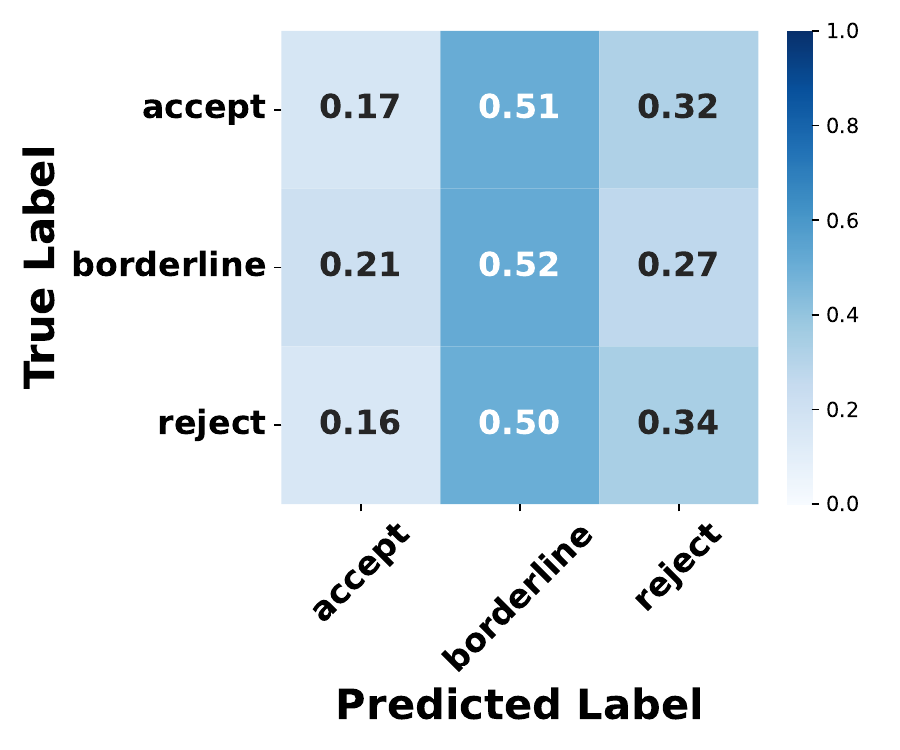}
        \caption{BERT Few-shot TBW ($\tau=0.3$)}
    \end{subfigure}
    \begin{subfigure}[b]{0.32\textwidth}
        \includegraphics[width=\textwidth]{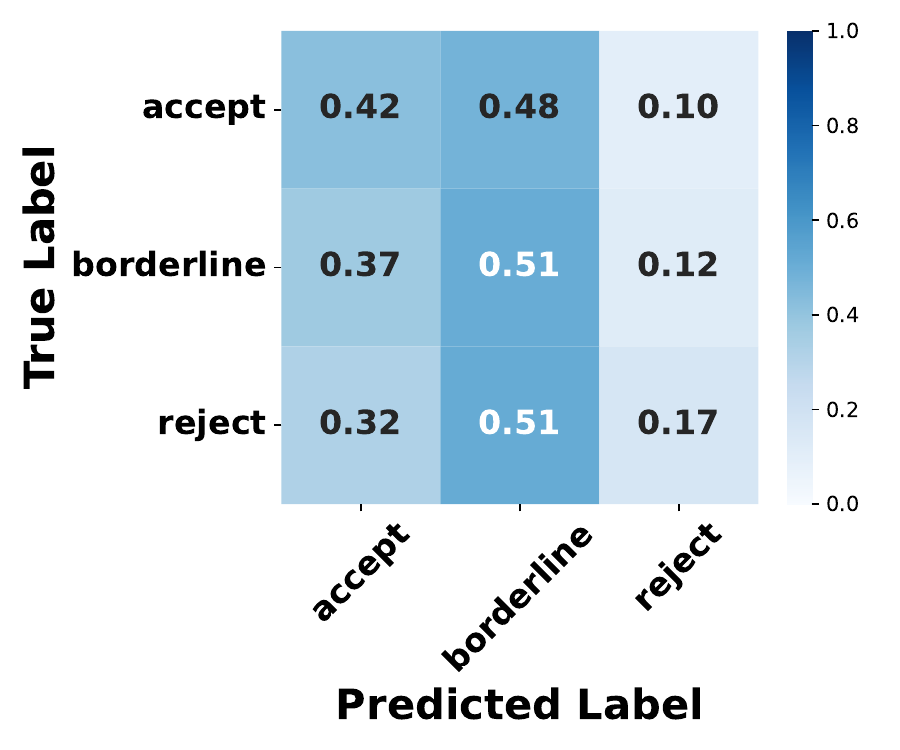}
        \caption{BERT Fine-tuned TBW ($\tau=0.3$)}
    \end{subfigure}
    
    \begin{subfigure}[b]{0.32\textwidth}
        \includegraphics[width=\textwidth]{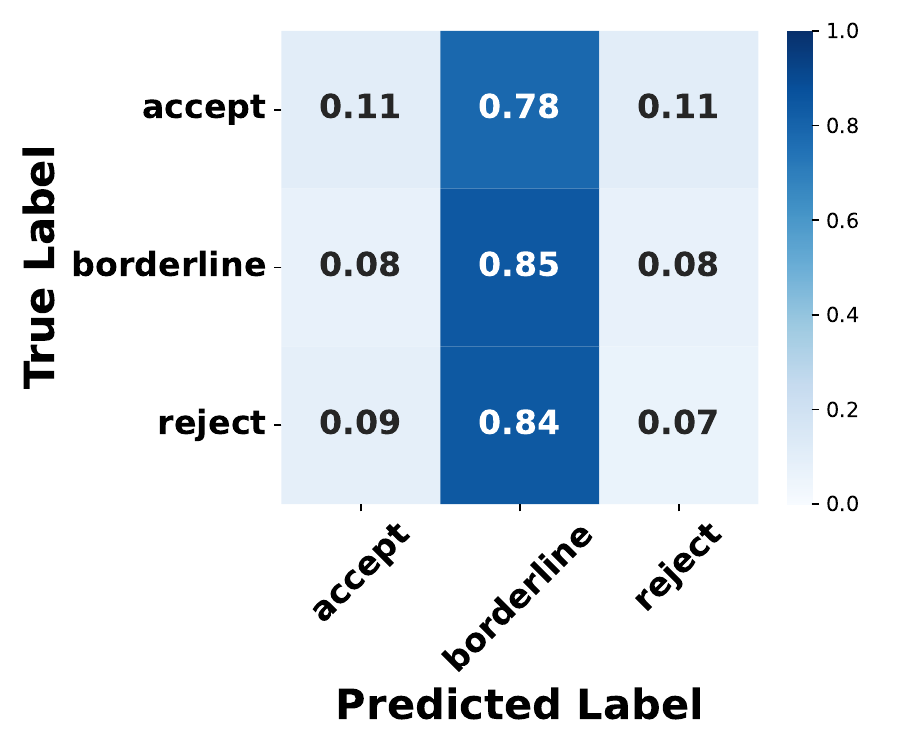}
        \caption{RoBERTa Base NW}
    \end{subfigure}
    \begin{subfigure}[b]{0.32\textwidth}
        \includegraphics[width=\textwidth]{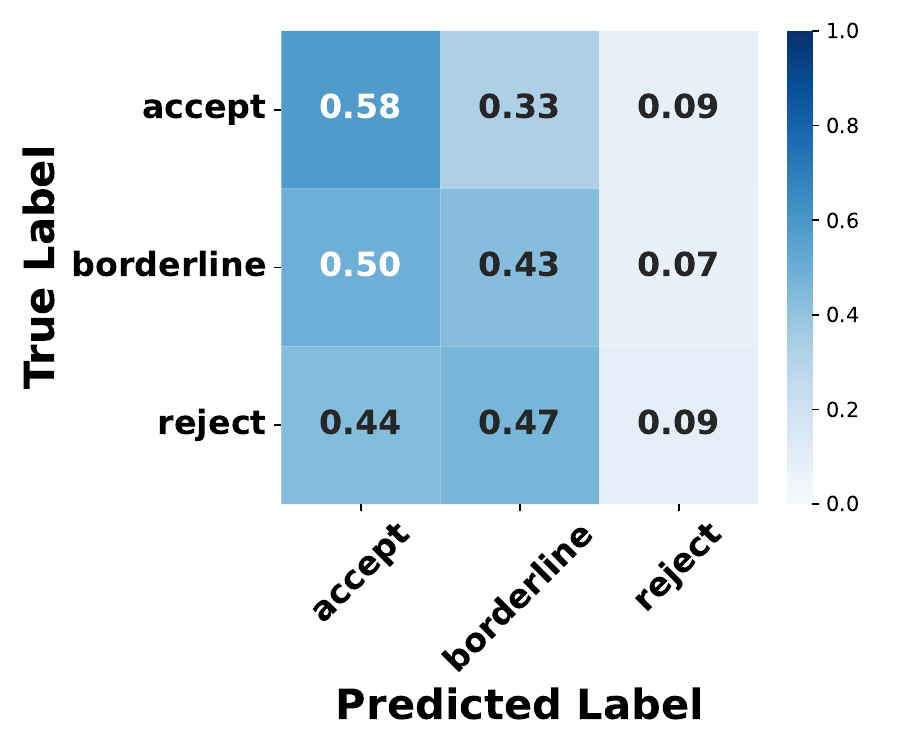}
        \caption{RoBERTa Few-shot NW}
    \end{subfigure}
    \begin{subfigure}[b]{0.32\textwidth}
        \includegraphics[width=\textwidth]{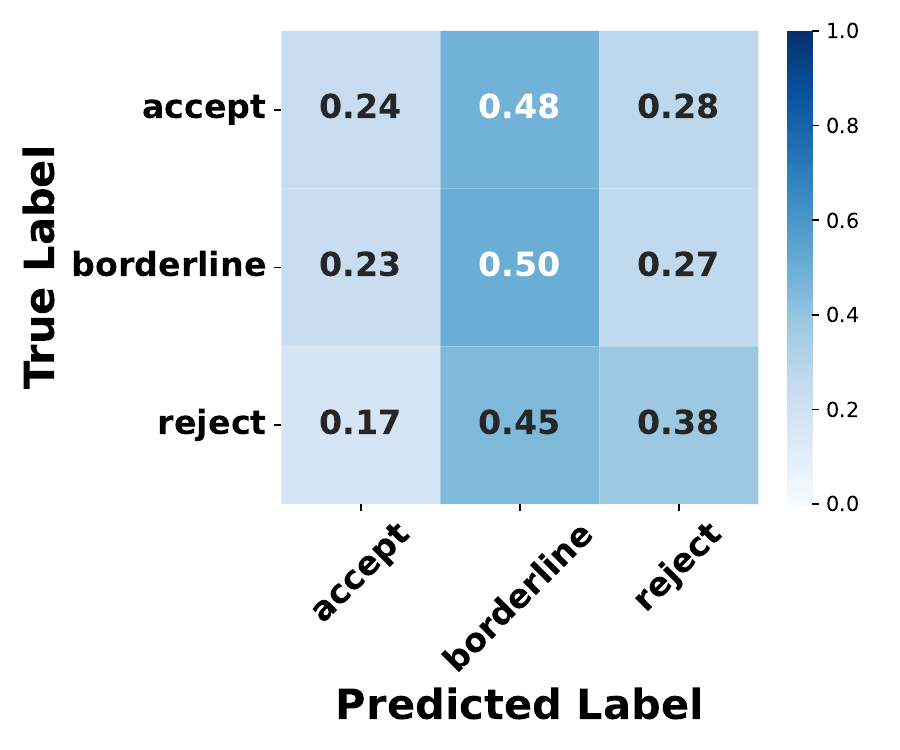}
        \caption{RoBERTa Fine-tuned NW}
    \end{subfigure}
    
    \begin{subfigure}[b]{0.32\textwidth}
        \includegraphics[width=\textwidth]{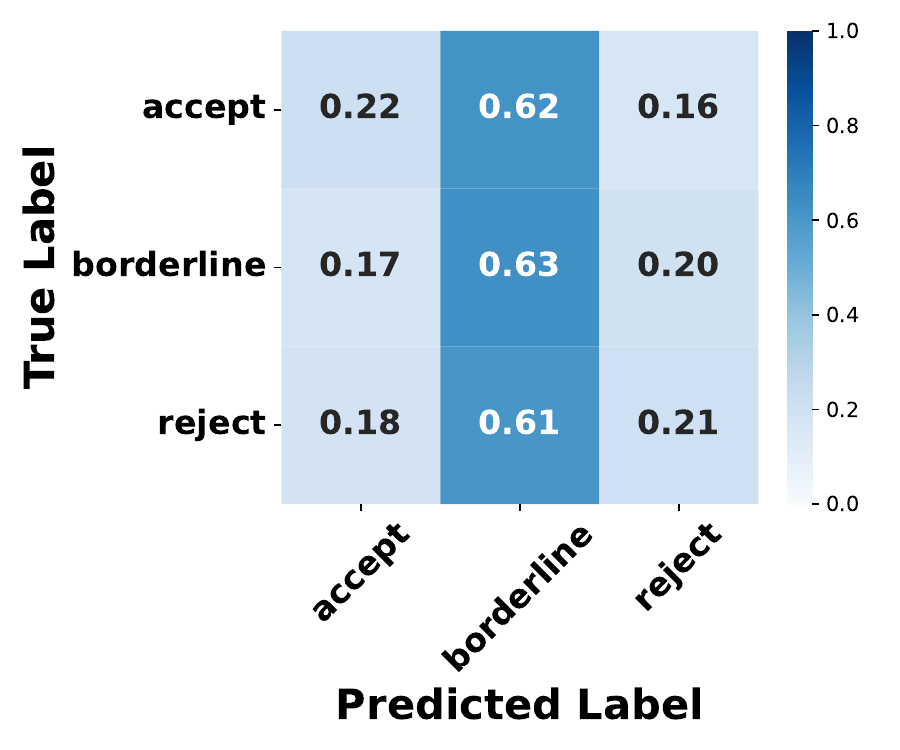}
        \caption{RoBERTa Base TBW ($\tau=0.7$)}
    \end{subfigure}
    \begin{subfigure}[b]{0.32\textwidth}
        \includegraphics[width=\textwidth]{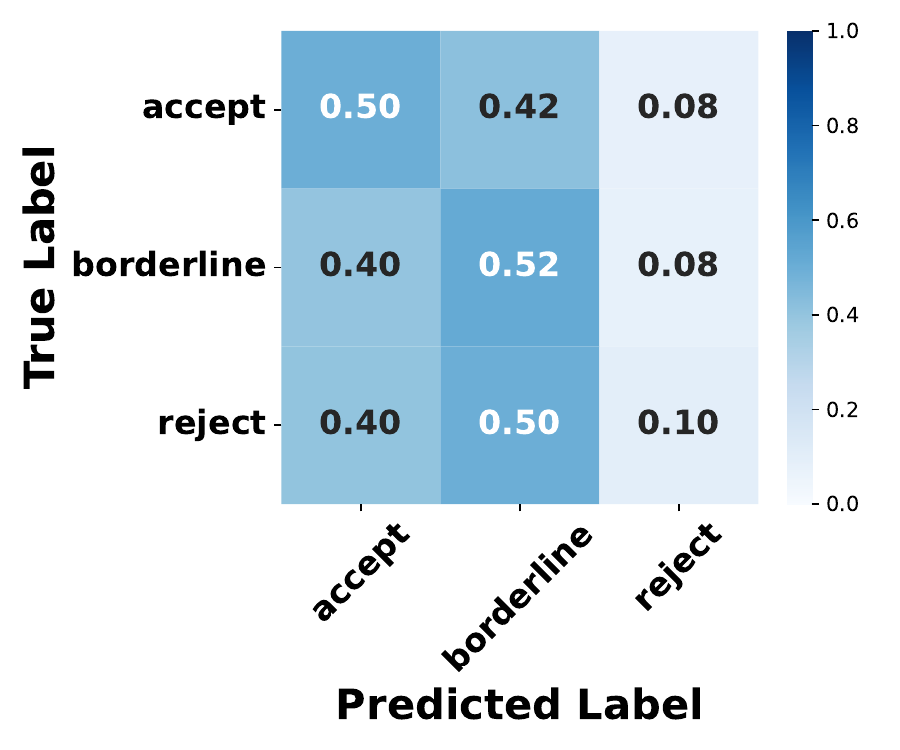}
        \caption{RoBERTa Few-shot TBW ($\tau=0.7$)}
    \end{subfigure}
    \begin{subfigure}[b]{0.32\textwidth}
        \includegraphics[width=\textwidth]{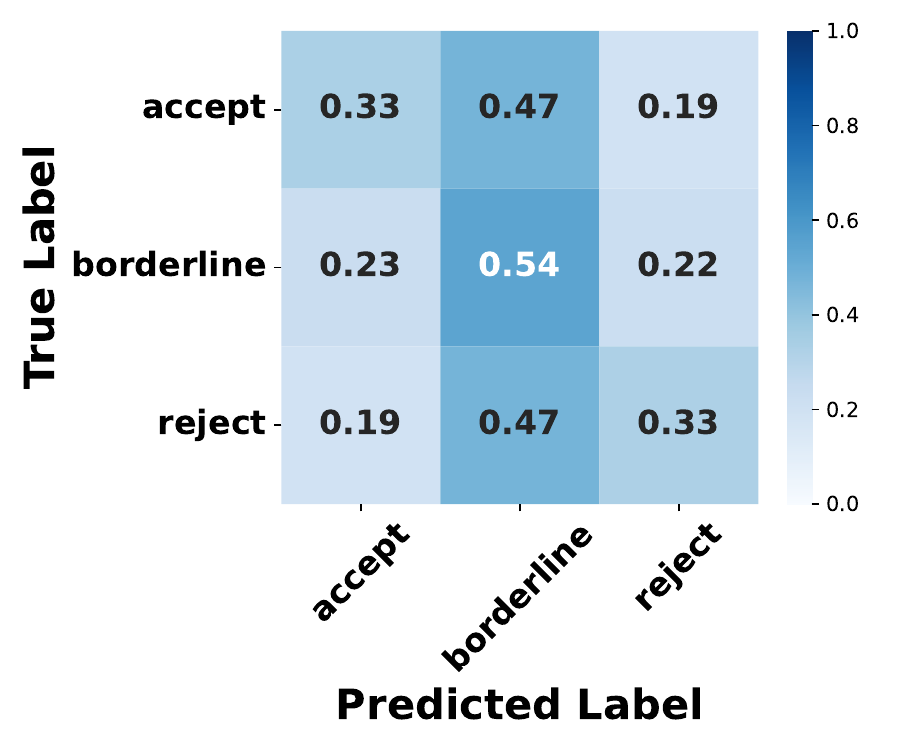}
        \caption{RoBERTa Fine-tuned TBW ($\tau=0.7$)}
    \end{subfigure}
    
    \begin{subfigure}[b]{0.32\textwidth}
        \includegraphics[width=\textwidth]{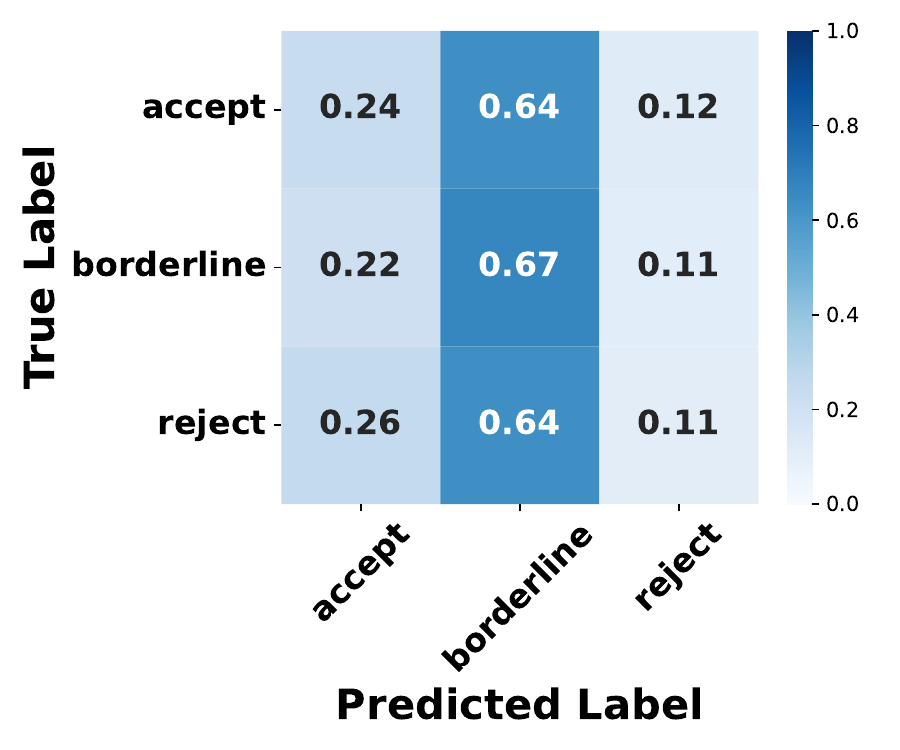}
        \caption{RoBERTa Base TBW ($\tau=0.3$)}
    \end{subfigure}
    \begin{subfigure}[b]{0.32\textwidth}
        \includegraphics[width=\textwidth]{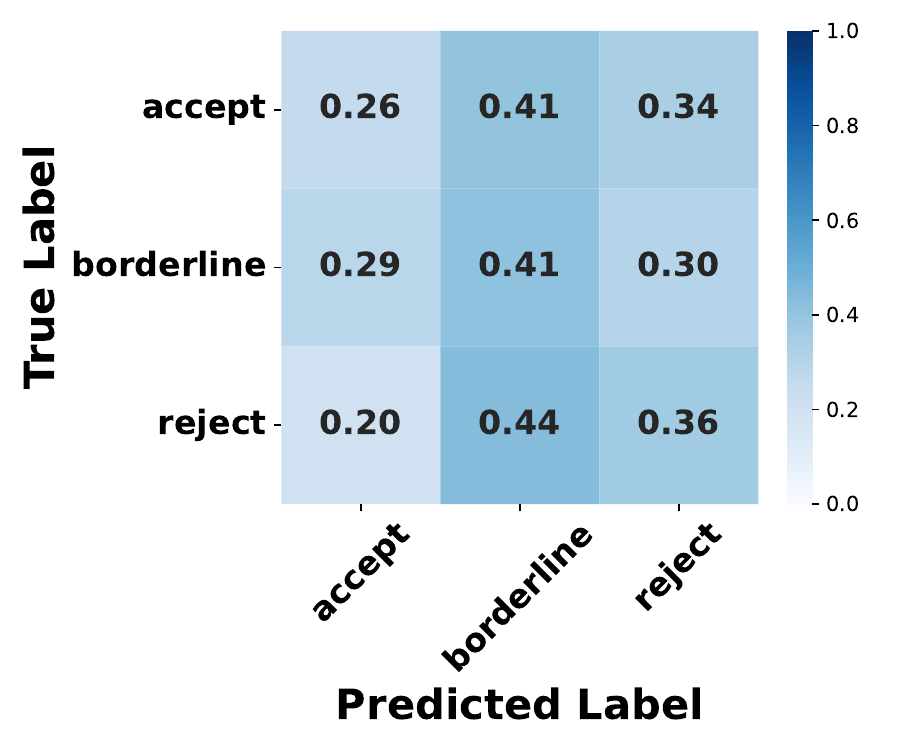}
        \caption{RoBERTa Few-shot TBW ($\tau=0.3$)}
    \end{subfigure}
    \begin{subfigure}[b]{0.32\textwidth}
        \includegraphics[width=\textwidth]{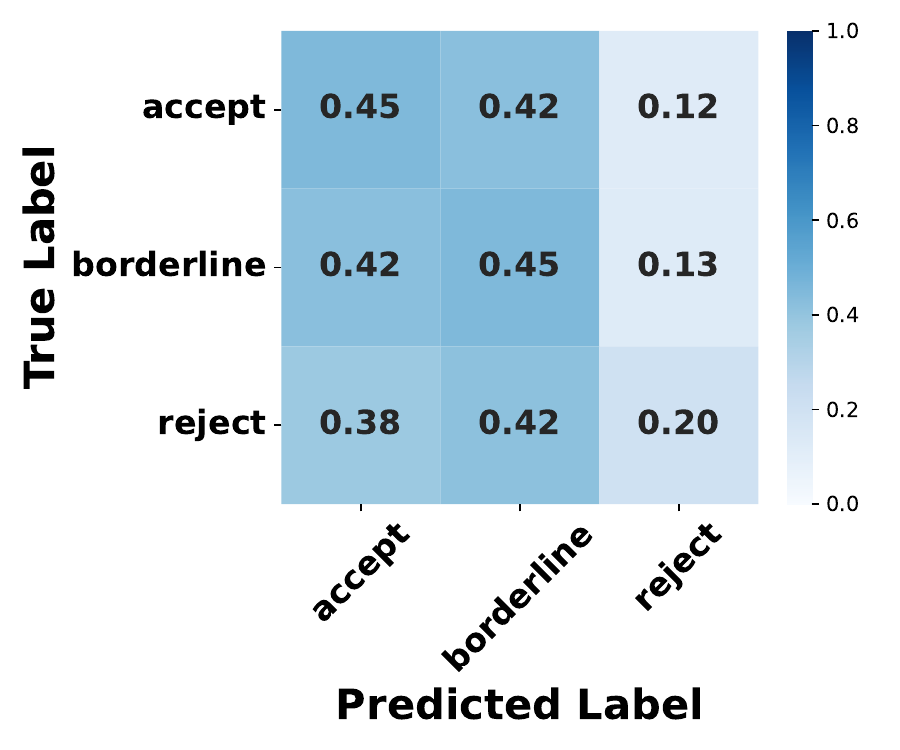}
        \caption{RoBERTa Fine-tuned TBW ($\tau=0.3$)}
    \end{subfigure}
    \end{minipage}
    }
    \caption{Comparison of confusion matrices across different model configurations and watermarking settings. Top row shows BERT results, middle row shows RoBERTa results, with columns representing no watermarking (NW), topic-based watermarking at $\tau = 0.7$, and topic-based watermarking at $\tau = 0.3$.}
    \label{fig:comprehensive_grid}
\end{figure*}

We examine confusion matrices for each classifier (BERT and RoBERTa), stratified by language model configuration (base, few-shot, fine-tuned) and watermarking condition. Additionally, we extend our analysis to topic-based watermarking (TBW) applied at a lower semantic similarity threshold of $\tau = 0.3$, which relaxes the token-to-topic alignment constraints, thereby increasing green-list coverage and watermark signal strength while potentially impacting semantic coherence. Figure~\ref{fig:comprehensive_grid} presents the complete set of confusion matrices across all configurations, including both $\tau = 0.7$ and $\tau = 0.3$ conditions.

BERT exhibits distinct patterns across model configurations and watermarking conditions. In the base non-watermarked condition (panel a), BERT shows a skew toward the \texttt{reject} column, while the few-shot variant (panel b) demonstrates higher predictions in the \texttt{accept} and \texttt{borderline} columns. The fine-tuned non-watermarked model (panel c) shows the highest concentration in the \texttt{borderline} column, though values remain below 0.50. Under watermarking conditions, BERT base models with both TBW $\tau = 0.7$ and $\tau = 0.3$ (panels d, g) exhibit a slight skew toward the \texttt{reject} column but with modest values barely exceeding 0.50. For few-shot and fine-tuned watermarked variants (panels e, f, h, i), predictions concentrate in the \texttt{borderline} column, with \texttt{accept} predictions consistently higher than \texttt{reject} but lower than \texttt{borderline}.

RoBERTa demonstrates more consistent patterns with clearer biases toward specific categories. Across all base configurations, non-watermarked, TBW $\tau = 0.7$, and TBW $\tau = 0.3$ (panels j, m, p), there is a pronounced bias toward the \texttt{borderline} column. The few-shot and fine-tuned variants generally show better-balanced distributions with higher concentrations in both \texttt{accept} and \texttt{borderline} columns. Notable exceptions include the RoBERTa few-shot TBW $\tau = 0.3$ condition (panel q), which maintains high \texttt{borderline} predictions, and the fine-tuned TBW $\tau = 0.7$ variant (panel o), which also shows elevated \texttt{borderline} concentrations.

\begin{table*}[t]
\centering
\caption{Classification performance for topic-based watermarking (TBW) at a lower similarity threshold of $\tau = 0.3$. Results are shown across all model configurations (base, few-shot, fine-tuned) and for both BERT and RoBERTa classifiers.}
\begin{tabular}{llcccc}
\toprule
\textbf{Classifier} & \textbf{Model} & \textbf{Accuracy} & \textbf{Precision} & \textbf{Recall} & \textbf{F1} \\
\midrule
\multirow{3}{*}{BERT} 
  & Base      & 0.289 & 0.322 & 0.322 & 0.288 \\
  & Few-shot  & 0.387 & 0.334 & 0.342 & 0.333 \\
  & Fine-tuned & 0.414 & 0.372 & 0.366 & 0.360 \\
\midrule
\multirow{3}{*}{RoBERTa} 
  & Base      & 0.438 & 0.338 & 0.340 & 0.332 \\
  & Few-shot  & 0.360 & 0.339 & 0.344 & 0.335 \\
  & Fine-tuned & 0.398 & 0.375 & 0.368 & 0.361 \\
\bottomrule
\end{tabular}
\label{tbw03-overall-perform}
\end{table*}

Table~\ref{tbw03-overall-perform} reports the classification metrics for each classifier and LLM model variant under TBW with $\tau = 0.3$. While overall performance remains comparable to the $\tau = 0.7$ condition, we observe that the fine-tuned model achieves the highest accuracy across both BERT and RoBERTa classifiers, suggesting that domain adaptation remains a dominant factor in attribution effectiveness even under relaxed topic alignment.
This analysis underscores the relative semantic distinctiveness of strongly positive (\texttt{accept}) and moderate (\texttt{borderline}) reviews, while highlighting the challenges involved in distinguishing lower-quality (\texttt{reject}) reviews, which often exhibit more linguistic and structural variability across different watermarking configurations and similarity thresholds.

\begin{figure}[t]
    \centering
    
    \begin{subfigure}[t]{0.48\textwidth}
        \centering
        \includegraphics[width=\textwidth]{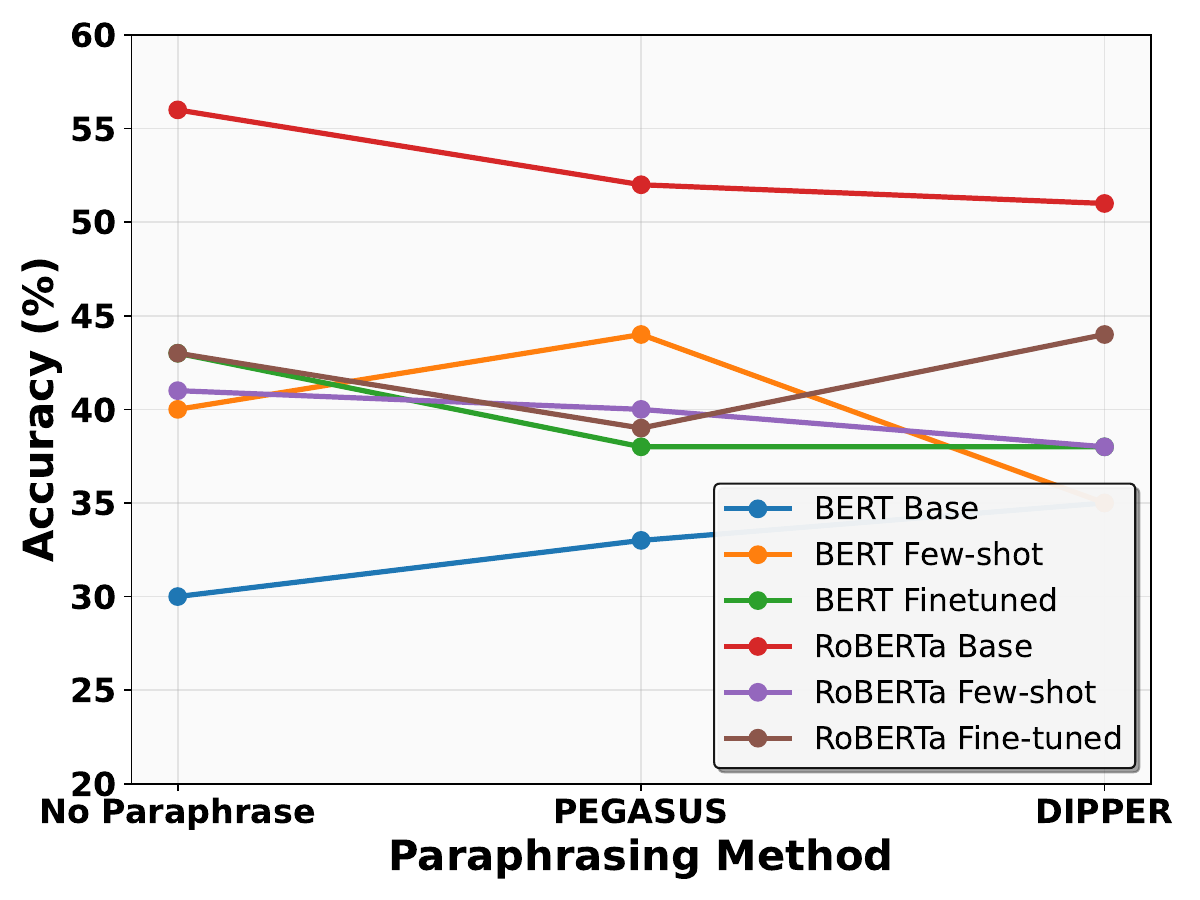}
        \caption{NW}
        \label{fig:fig1}
    \end{subfigure}
    
    \vspace{0.5cm}
    
    \begin{subfigure}[t]{0.48\textwidth}
        \centering
        \includegraphics[width=\textwidth]{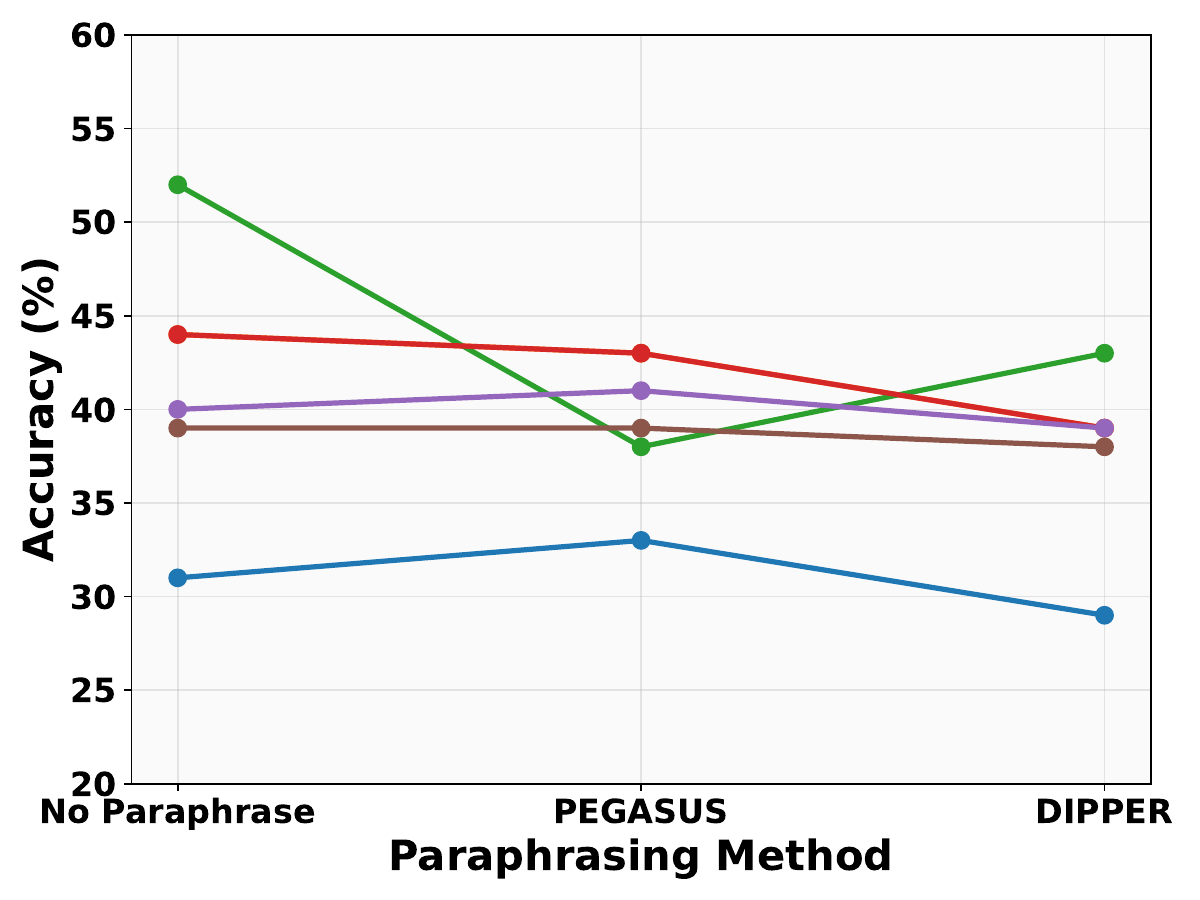}
        \caption{TBW $\tau=0.7$}
        \label{fig:fig2}
    \end{subfigure}
    
    \vspace{0.5cm}
    
    \begin{subfigure}[t]{0.48\textwidth}
        \centering
        \includegraphics[width=\textwidth]{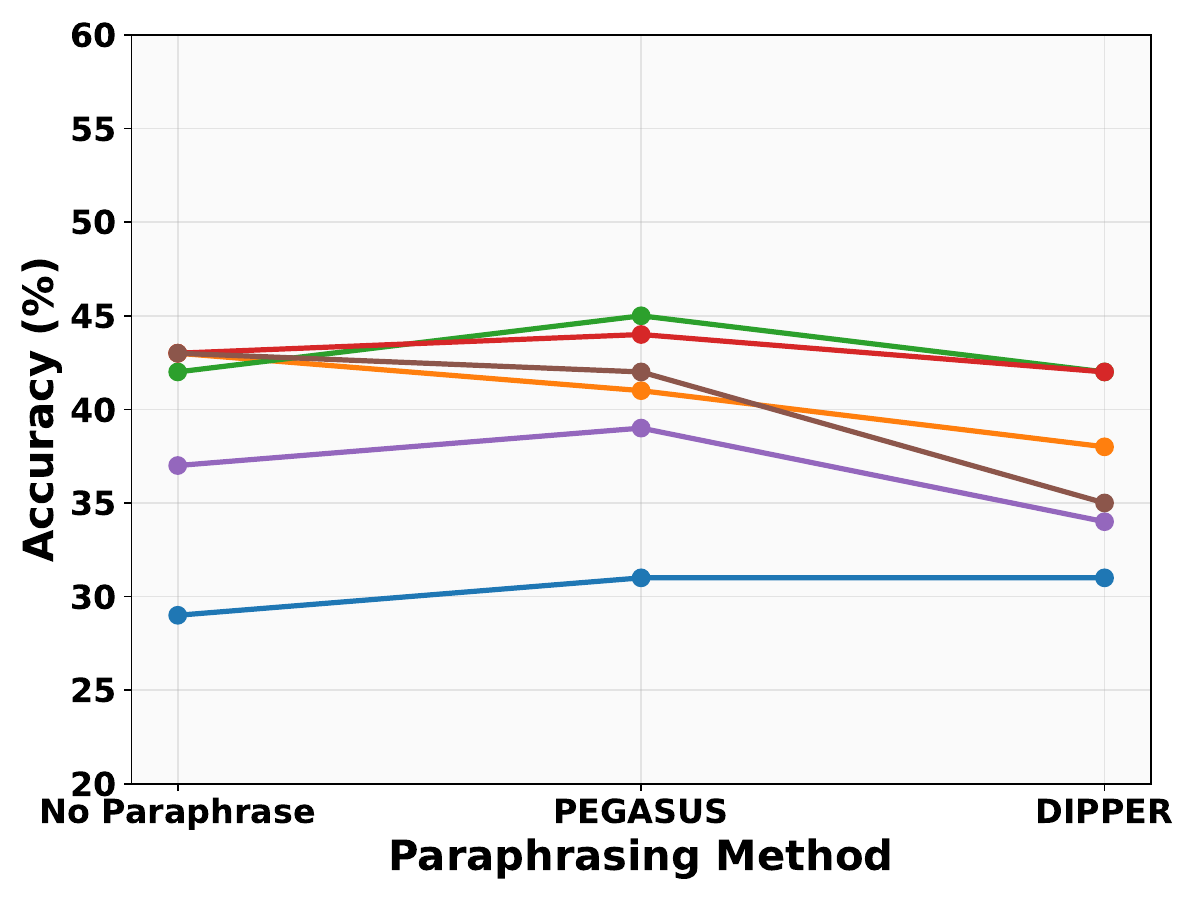}
        \caption{TBW $\tau=0.3$}
        \label{fig:fig3}
    \end{subfigure}
    
\caption{Classification accuracy on paraphrased peer reviews across three watermarking settings: (a) no watermark (NW), (b) topic-based watermarking (TBW) with $\tau = 0.7$, and (c) TBW with $\tau = 0.3$. Results are shown across all model configurations (base, few-shot, fine-tuned) for both BERT and RoBERTa classifiers under PEGASUS and DIPPER paraphrasing attacks.}
    \label{fig:three_figures}
\end{figure}

\subsection{Peer Review Shifts Under Paraphrasing}\label{peerreviewparaphrase}
To evaluate the impact of paraphrasing on classifier-based review attribution, we examine both classification accuracy and label stability under two paraphrasing threat models: PEGASUS and DIPPER. Specifically, we sample 100 LLM-generated peer reviews and apply paraphrasing to each using both models. We then assess the classification performance before and after paraphrasing under three watermarking conditions: no watermark (NW), topic-based watermarking (TBW) with $\tau = 0.7$, and TBW with $\tau = 0.3$.

Figure~\ref{fig:three_figures} presents accuracy changes across all classifier and model configurations. Table~\ref{shifts} reports the number of label transitions (e.g., \texttt{Accept} $\rightarrow$ \texttt{Borderline}) observed in the paraphrased reviews. These metrics reflect the semantic resilience of reviewer intent and classification stability under adversarial rewording.

Our results indicate that paraphrasing generally reduces classification accuracy across all settings, though the degree of degradation varies. Notably, TBW models exhibit consistent accuracy declines under paraphrasing for both $\tau$ values, suggesting that watermarked outputs are more sensitive to adversarial modification in terms of downstream attribution. In contrast, non-watermarked outputs show mixed effects while some configurations experience accuracy drops, others see minor improvements. We attribute this to incidental lexical clarifications introduced by the paraphrasers. In terms of label stability, TBW reduces the number of class shifts compared to the non-watermarked baseline. This trend is especially evident under the PEGASUS paraphrasing model, where non-watermarked outputs exhibit the highest number of shifts. These findings suggest that TBW not only leaves a detectable signature but may also provide a degree of structural regularity that preserves classification under text manipulation.

\begin{table}[h!]
\centering
\scriptsize
\begin{tabular}{@{}lllcc@{}}
\toprule
\textbf{Classifier} & \textbf{Model} & \textbf{Watermark} & \textbf{PEGASUS} & \textbf{DIPPER} \\
& & & \textbf{Shifts} & \textbf{Shifts} \\
\midrule
\multirow{9}{*}{\textbf{BERT}} 
  & \multirow{3}{*}{Base}     
    & No watermark    & 58 & 54 \\
  &             & TBW ($\tau=0.7$) & 37 & 23 \\
  &             & TBW ($\tau=0.3$) & 51 & 45 \\
  \cmidrule(lr){2-5}
  & \multirow{3}{*}{Few-shot} 
    & No watermark    & 24 & 14 \\
  &             & TBW ($\tau=0.7$) & 24 & 24 \\
  &             & TBW ($\tau=0.3$) & 24 & 22 \\
  \cmidrule(lr){2-5}
  & \multirow{3}{*}{Fine-tuned}
    & No watermark    & 27 & 20 \\
  &             & TBW ($\tau=0.7$) & 15 & 15 \\
  &             & TBW ($\tau=0.3$) & 25 & 15 \\
\midrule
\multirow{9}{*}{\textbf{RoBERTa}} 
  & \multirow{3}{*}{Base}     
    & No watermark    & 13 & 9 \\
  &             & TBW ($\tau=0.7$) & 23 & 25 \\
  &             & TBW ($\tau=0.3$) & 16 & 19 \\
  \cmidrule(lr){2-5}
  & \multirow{3}{*}{Few-shot} 
    & No watermark    & 30 & 13 \\
  &             & TBW ($\tau=0.7$) & 27 & 22 \\
  &             & TBW ($\tau=0.3$) & 25 & 20 \\
  \cmidrule(lr){2-5}
  & \multirow{3}{*}{Fine-tuned}
    & No watermark    & 24 & 14 \\
  &             & TBW ($\tau=0.7$) & 18 & 22 \\
  &             & TBW ($\tau=0.3$) & 21 & 18 \\
\bottomrule
\end{tabular}
\caption{Number of review classification shifts under paraphrasing attacks. Each entry reflects the count (out of 100 paraphrased samples) where the predicted class label differs from the original. Results are grouped by classifier, model variant, and watermarking scheme (NW, TBW ($\tau=0.7$), TBW ($\tau=0.3$)), and evaluated separately under PEGASUS and DIPPER paraphrasing models.}
\label{shifts}
\end{table}

\end{document}